\documentclass[amsmath,amssymb,aps,nofootinbib,nolongbibliography]{revtex4-2}
\usepackage{bm}
\usepackage{dcolumn}
\usepackage{graphicx}
\usepackage{multirow}
\usepackage{subcaption}
\usepackage{color}
\usepackage{soul}

\begin{document}

\title{Revisiting series expansions of neutrino oscillation and decay probabilities in matter}

\author{Jesper Gr\"onroos}
\author{Tommy Ohlsson}
\author{Sampsa Vihonen}

\affiliation{Department of Physics, School of Engineering Sciences, KTH Royal Institute of Technology,\\
AlbaNova University Center, Roslagstullsbacken 21, SE–106 91 Stockholm, Sweden}
\affiliation{The Oskar Klein Centre, AlbaNova University Center, Roslagstullsbacken 21,\\
SE–106 91 Stockholm, Sweden}

\email{jesgro@kth.se}
\email{tohlsson@kth.se}
\email{vihonen@kth.se}

\date{\today}

\begin{abstract}
We present analytic expressions for three-flavor neutrino oscillations in presence of invisible neutrino decay and matter effects. Using the well-known Cayley--Hamilton formalism, the leading-order terms are derived for oscillation probabilities in all major channels assuming the neutrino mass eigenstate $\nu_3$ to decay. Our work extends and complements previous studies utilizing the Cayley--Hamilton theorem, providing the series expansions for $\nu_e \rightarrow \nu_e$, $\nu_e \rightarrow \nu_\mu$, $\nu_e \rightarrow \nu_\tau$, $\nu_\mu \rightarrow \nu_\mu$ and $\nu_\mu \rightarrow \nu_\tau$. The accuracy of the analytical formulas is investigated by comparing the results with numerically calculated probabilities. We also comment on the implications on unitarity violation.
\end{abstract}

\maketitle
%\flushbottom

\section{Introduction}
\label{sec:intro}
Neutrino decay was one of the alternative solutions that were proposed to describe the observed zenith angle dependence on the early atmospheric neutrino data in Super-Kamiokande in 1998~\cite{Barger:1998xk}. According to the decay paradigm, neutrinos can decay into lighter states in order to achieve the muon neutrino disappearance. In such case, the decay product could be either visible or invisible. If neutrinos were to be Dirac particles and the decay was the invisible kind, the decay product could consist of a right-handed $SU(2)$ singlet and a complex scalar $\chi$ with lepton number $-2$ and weak isospin and hypercharge both equal to zero~\cite{Acker:1991ej,Acker:1993sz}. In the case neutrinos were Majorana particles, the decay product would include a Majoron $J$ instead~\cite{Gelmini:1980re,Chikashige:1980ui}. In both cases, the decay product would be invisible to the neutrino detector. While the attention on neutrino decay has declined as a description of the solar neutrino deficit, there are still models that predict both visible and invisible neutrino decay, see {\em e.g.} Refs.~\cite{Cogollo:2008zc,Cogollo:2012ek,Picoreti:2015ika,SNO:2018pvg,Hostert:2020oui,Picoreti:2021yct,deGouvea:2022cmo,deGouvea:2023jxn,Martinez-Mirave:2024hfd}. In such cases, neutrino decay leads to sub-leading corrections that could be probed in future neutrino experiments.

Experimental research on neutrino oscillation physics has witnessed more than two decades of uninterrupted success. After the discovery of neutrino oscillations, numerous experiments probing neutrino oscillations with reactor, solar, atmospheric and accelerator origin have maintained that neutrino oscillations can be described with the standard oscillation parameters, three mixing angles $\theta_{12}$, $\theta_{13}$ and $\theta_{23}$, two mass-squared differences $\Delta m_{21}^2$ and $\Delta m_{31}^2$, and the Dirac \emph{CP} phase $\delta$. The remaining unknowns of the standard parameters include the sign of $\Delta m_{31}^2$, which can follow either normal neutrino mass ordering, $\Delta m_{31}^2 >$ 0, or inverted ordering, $\Delta m_{31}^2 <$ 0. Also unknown is the true value of $\delta$, which may be \emph{CP}-violating, $\sin \delta \neq$ 0, or \emph{CP}-conserving, $\sin \delta =$ 0. The increased precision on the standard neutrino mixing parameters has also led to stringent bounds on beyond the Standard Model physics models that predict neutrino decay. The most constraining bounds on the invisible decay of the neutrino mass states $m_1$ and $m_2$ arise from the combined analysis of solar neutrino data and KamLAND reactor neutrino data~\cite{Berryman:2014qha}, $\tau_1/m_1 >$ 4$\times$10$^{-3}$~s/eV and  $\tau_2/m_2 >$ 7$\times$10$^{-4}$~s/eV (95\% C.L.), where $\tau_i$ denotes the lifetime of mass state $\nu_i$, $i=$ 1, 2, 3. Global analysis on long-baseline and atmospheric neutrino data~\cite{Gonzalez-Garcia:2008mgl} has set furthermore a lower bound on the heaviest mass state with $\tau_3/m_3 >$ 2.9$\times$10$^{-10}$~s/eV. Regarding the constraints on complete neutrino decay, supernova neutrino data from SN1987 sets lower bound on electron neutrino lifetime $\tau_{\nu_e} >$ 5.0$\times$10$^{5} (m_{\nu_e} /\text{eV})$~\cite{Frieman:1987as}, which translates into $\tau_i/m_i \geq$ 1.2$\times$10$^{5}$~s/eV for mass states $m_1$ and $m_2$ at 90\% CL~\cite{Ivanez-Ballesteros:2023lqa}. With the coming of the next-generation neutrino telescopes like IceCube Gen-2~\cite{IceCube-Gen2:2020qha} and KM3NeT~\cite{KM3Net:2016zxf} and laboratory experiments like DUNE~\cite{DUNE:2015lol}, ESSnuSB~\cite{ESSnuSB:2023ogw} and HyperKamiokande~\cite{Abe:2015zbg}, the limits on invisible neutrino decay are expected to improve.

Approximating neutrino oscillation probabilities with neutrino decay is a non-trivial matter. In the most general formalization of neutrino decay, the effective Hamiltonian consists of Hermitian and anti-Hermitian parts, the latter of which accounts for neutrino decay. As the eigenvalues of the Hermitian and anti-Hermitian parts do not necessarily match, the two matrices must be diagonalized independently. Compact analytic expressions for three-flavor oscillations were derived in Ref.~\cite{Lindner:2001fx} and reviewed in Refs.~\cite{Abrahao:2015rba,Ghoshal:2020hyo}. Compact analytic forms were first obtained for non-commuting Hermitian and anti-Hermitian operators in two neutrino flavors~\cite{Chattopadhyay:2021eba} and later for three flavors~\cite{Chattopadhyay:2022ftv} using Zassenhaus expansions~\cite{Kimura:2017xxz}. The neutrino oscillation probabilities, which we normally denote as $P_{\alpha\beta}$ for $\nu_\alpha$ oscillating into $\nu_\beta$ ($\alpha, \beta = e, \mu, \tau$), are currently known for $P_{ee}$ up to order $\mathcal{O}(\lambda^2)$ and for $P_{e\mu}$ to $\mathcal{O}(\lambda^3)$, where $\lambda \equiv$ 0.2. The survival probability $P_{\mu\mu}$ has also been derived to zeroth order.

In the present work, we provide the full set of neutrino oscillation probabilities in the scenario where the heaviest neutrino mass state is able to decay. Employing the method introduced in Ref.~\cite{Akhmedov:2004ny}, we present series expansions for the neutrino oscillation probabilities $P_{ee}$, $P_{e\mu}$, $P_{e\tau}$, $P_{\mu \mu}$ and $P_{\mu \tau}$ in terms of the small parameters $\alpha \equiv \Delta m_{21}^2 / \Delta m_{31}^2$ and $s_{13} \equiv \sin \theta_{13}$. Matter effects are taken into account throughout this work. The analytical expressions presented are given for $P_{ee}$ up to order $\mathcal{O}(\lambda^2)$, and for $P_{e\mu}$, $P_{e\tau}$, $P_{\mu \mu}$ and $P_{\mu \tau}$ to order $\mathcal{O}(\lambda^3)$, respectively. While the results derived for $P_{ee}$ and $P_{e\mu}$ are in agreement with those presented in Ref.~\cite{Chattopadhyay:2022ftv}, the formulas derived for $P_{\mu\mu}$ and $P_{\mu\tau}$ are novel results.

This article is organized as follows. We review the phenomenology of invisible neutrino decay in section~\ref{sec:background}, discussing both generalized decay where all neutrino mass states can decay and also the special case where only the heaviest neutrino mass state decays. The perturbative methods that are used in this work are also briefly described. Compact analytical formulas for oscillation probabilities with neutrino decay are presented in section~\ref{sec:expansions}, giving a full account on matter effects arising from neutrino propagation in matter. The oscillation probabilities are also obtained in the vacuum limit in this section. We moreover present the unitarity relations in presence of neutrino decay. The applications and limitations of these formulas are discussed in section~\ref{Discussion} and concluding remarks are presented in section~\ref{sec:summary}. The Cayley--Hamilton formalism and perturbative diagonalization are described in appendices~\ref{sec:appA} and \ref{sec:appB}, respectively.

\section{\label{sec:background}Methodology}

The aim of this work is to obtain compact analytical expressions for neutrino oscillation probabilities in presence of neutrino decay and matter effects. To achieve this, we compute the time-evolution operator $S = \text{e}^{-i\mathcal{H}L}$ for effective decay Hamiltonian $\mathcal{H}$ and neutrino propagation length $L$ and obtain probabilities for $\nu_\alpha \rightarrow \nu_\beta$ as $P_{\alpha\beta}=|S_{\beta\alpha}|^2$. In this section, we review the effective decay Hamiltonian for general decay matrix $\Gamma$ and also for the case where one active neutrino decays. We also describe the perturbative methods that are used to obtain the compact analytical formulas.

\subsection{\label{sec:background:formalism}Effective decay Hamiltonian}

Neutrino decay can be described with the effective Hamiltonian that consists of Hermitian and anti-Hermitian parts:
\begin{equation}
\label{eq:01}
\mathcal{H}_m = H_m - \frac{i}{2} \Gamma_m.
\end{equation}
The Hermitian $H_m$ is defined in the mass basis and contains the dynamics of standard neutrino oscillations. The anti-Hermitian component $\Gamma_m$ on the other hand is a $3\times3$ matrix that contains the elements leading to neutrino decay. In this case, the decay of neutrino mass state $\nu_i$ is best characterized by the term exp[$(-m_i/\tau_i)/(L/E)$], where $\tau_i$ represents the lifetime while $L$ and $E$, respectively, give the propagation distance and energy of the decaying neutrino state $\nu_i$. 

The anti-Hermitian component in equation~(\ref{eq:01}) can be written in Weisskopf--Wigner approximation:
\begin{equation}
\label{eq:02}
\Gamma_{ij} = 2\pi \sum_k \langle \nu_i | \mathcal{H}' | \phi_k \rangle \langle \phi_k | \mathcal{H}' | \nu_j \rangle \, \delta (E_k - E),
\end{equation}
where $|\phi_k\rangle$ is the final state representing decay products with energy $E_k$. The interaction term $\mathcal{H}'$ is responsible for neutrino decay and it arises from beyond the Standard Model physics.

The difficulty of obtaining compact formulas with the effective Hamiltonian $\mathcal{H}_m$  arises from the anti-Hermitian decay component $\Gamma_m$ in equation~(\ref{eq:01}). Since neutrino mass and decay eigenstates do not necessarily need to coincide, the Hermitian and anti-Hermitian components $H_m$ and $\Gamma_m$ do not always commute and neutrino oscillation probabilities must therefore be expressed in terms of $[H_m, \Gamma_m] \neq$ 0. Neutrino interactions with matter further complicate the calculation, as vacuum eigenstates of the Hamiltonian are generally different from the eigenstates in matter.

When neutrinos are propagating in matter, the generalized Hamiltonian can be written as follows
\begin{equation}
\label{eq:03}
\mathcal{H}_f (\Gamma) = U \left[ \frac{1}{2 E} \begin{pmatrix}0 & 0 & 0\\0 & \Delta m_{21}^2 & 0\\0 & 0 & \Delta m_{31}^2\end{pmatrix} - \frac{i}{2} \Gamma \right]U^\dagger + \begin{pmatrix}V_{CC} & 0 & 0\\0 & 0 & 0\\0 & 0 & 0\end{pmatrix}.
\end{equation}
Here $\mathcal{H}_f$ is given in flavor basis, $U$ is the Pontecorvo--Maki--Nakagawa--Sakata matrix~\cite{Pontecorvo:1957cp,Maki:1962mu} and $V_{CC}=\sqrt{2} G_F N_e$ is the standard matter potential consisting of Fermi coupling constant $G_f$ and electron density $N_e$, respectively.

It is often more convenient to focus on a special case where only one neutrino mass state undergoes invisible decay. In the case of normal neutrino mass ordering, the decaying mass state would be $\nu_3$. If the mass ordering was inverted, the decaying mass state would be $\nu_2$ instead. In the case it is $\nu_3$ that decays, it is enough to take one decay parameter $\gamma$ and give effective Hamiltonian as follows \begin{equation}
\label{eq:04}
\mathcal{H}_f (\gamma) = \frac{\Delta m_{31}^2}{2 E} U \left[\begin{pmatrix}0 & 0 & 0\\0 & \alpha & 0\\0 & 0 & 1\end{pmatrix} - i \begin{pmatrix}0 & 0 & 0\\0 & 0 & 0\\0 & 0 & \gamma\end{pmatrix}\right]U^\dagger + \begin{pmatrix}V_{CC} & 0 & 0\\0 & 0 & 0\\0 & 0 & 0\end{pmatrix},
\end{equation}
where $\alpha \equiv \Delta m_{21}^2 / \Delta m_{31}^2$ is a small parameter. The decay parameter $\gamma$ can then be written as $\gamma \equiv m_3/(\tau_3 \Delta m_{31}^2)$ and it is always real. It is also possible to define the effective Hamiltonian for $\nu_2$ decay.

\subsection{Time-evolution operator}
\label{Method-1}

Neutrino oscillation probabilities are generally calculated with the time-evolution operator $S \equiv \text{e}^{-i \mathcal{H}_f L}$, where $\mathcal{H}_f = \mathcal{H}_f (\gamma)$ in equation~(\ref{eq:04}) is the effective Hamiltonian and $L$ is the propagation distance of neutrinos. Probabilities for the channels $\nu_\alpha \rightarrow \nu_\beta$ ($\alpha$, $\beta = e$, $\mu$, $\tau$) are then obtained as the square of the absolute value of the time-evolution operator, that is, $P_{\alpha \beta} = \left| S_{\beta \alpha} \right|^2$. We derive analytical probabilities by calculating series expansions in terms of $s_{13}$ and $\alpha$. Using the Cayley--Hamilton formalism in general, such expansions can in principle be carried out for $N$ neutrino flavors~\cite{Ohlsson:1999xb, Ohlsson:1999um, Ohlsson:2001vp}. In this work, perturbative diagonalization is used to obtain the eigenvalues of the effective Hamiltonian. The eigenvalues and eigenvectors are then used to obtain analytical formulas for neutrino oscillation probabilities. We carry out the derivation of the series expansions in \verb|Mathematica|. For convenience, the results are grouped in terms of the book-keeping parameter $\lambda \equiv 0.2$, which allows to express the expansion parameters as $s_{13} \simeq 0.14 \sim \mathcal{O}(\lambda)$ and $\alpha \simeq 0.03 \sim \mathcal{O}(\lambda^2)$. We provide more information on the Cayley--Hamilton formalism in appendix~\ref{sec:appA}, whereas the perturbative diagonalization method is summarized in appendix~\ref{sec:appB}.

Working in the flavor basis, using the Cayley--Hamiltonian formalism, the time-evolution operator for neutrino propagation can be written as (see {\em e.g.}~Ref.~\cite{Akhmedov:2004ny})
\begin{equation}
    \label{timeevolution}
    \begin{split}
    S = \text{e}^{-i{\mathcal{H} L}} &= \frac{\text{e}^{-i{ E_1}L}}{(E_1-E_2)(E_1-E_3)}\Big[E_2 E_3\boldsymbol{I}- (E_2+E_3)\mathcal{H} + \mathcal{H}^2 \Big]\\ 
    &+\frac{\text{e}^{-i E_2L}}{(E_2-E_3)(E_2-E_1)}\Big[E_3 E_1\boldsymbol{I}- (E_3+E_1)\mathcal{H} + \mathcal{H}^2 \Big]\\
    &+\frac{\text{e}^{-i E_3L}}{(E_3-E_1)(E_3-E_2)}\Big[E_1 E_2\boldsymbol{I}- (E_1+E_2)\mathcal{H} + \mathcal{H}^2 \Big],
    \end{split}
\end{equation}
where $E_i$ are the eigenvalues of the Hamiltonian and $\boldsymbol{I}$ is the identity matrix. For three-flavor oscillations, $S$ is a complex $3\times 3$-matrix. We derive the energy eigenvalues of equation~(\ref{timeevolution}) up to first order in $s_{13}$ and $\alpha$:
\begin{align}
        E_1 &\simeq \frac{\Delta m_{31}^2}{2E}\bigg(A+\alpha s_{12}^2+s_{13}^2\frac{A(1-i \gamma)}{A-1+i \gamma}\bigg),\label{energyeigenvalues1}\\
        E_2 &\simeq \frac{\Delta m_{31}^2}{2E} \alpha c_{12}^2,\label{energyeigenvalues2}\\
        E_3 &\simeq \frac{\Delta m_{31}^2}{2E}\bigg(1-i \gamma -s_{13}^2\frac{A(1-i\gamma)}{A-1+i\gamma}\bigg),\label{energyeigenvalues3}
\end{align}
where $A \equiv 2V_{CC} E / \Delta m_{31}^2$ is the standard matter potential arising from neutrino interactions in matter and $\gamma$ is the decay parameter in the simplest scenario where the heaviest neutrino decays. We find the eigenvalues presented in equations~(\ref{energyeigenvalues1})--(\ref{energyeigenvalues3}) to be consistent with those those shown in Ref.~\cite{Chattopadhyay:2022ftv}. We are also able to recover the eigenvalues derived for the standard oscillations in Ref.~\cite{Akhmedov:2004ny} by applying $\gamma \rightarrow 0$ and $A \rightarrow 0$.

The series expansions derived in this work are grouped in terms that are proportional to various powers of $\lambda$. In this arrangement, terms $P_{\alpha\beta}^{(0)}$ are independent of $s_{13}$ and $\alpha$ and they therefore correspond to $\mathcal{O}(\lambda^0) = \mathcal{O}(1)$. On the other hand, terms $P_{\alpha\beta}^{(1)}$ are proportional to $s_{13}$ and correspond to $\mathcal{O}(\lambda)$, whereas terms $P_{\alpha\beta}^{(2)}$ are proportional to either $s_{13}^2$ or $\alpha$ and correspond to $\mathcal{O}(\lambda^2)$. Similarly, terms $P_{\alpha\beta}^{(3)}$ that are proportional to either $s_{13}^3$ or $\alpha s_{13}$ and are associated with $\mathcal{O}(\lambda^3)$. Probabilities involving $\nu_e$ can be obtained completely and they are presented in Section~\ref{sec:expansions}. Probabilities involving only $\nu_\mu$ and $\nu_\tau$ are more complicated to derive than those involving $\nu_e$. In those cases, the expressions become so large that it is more beneficial to compute them as sum of terms to all relevant orders of parameter $\lambda$. As such, we write probabilities involving only $\nu_\mu$ and $\nu_\tau$ as
\begin{equation}
\label{series_expansion}
    P_{\alpha\beta} = P_{\alpha\beta}^{(0)} +  P_{\alpha\beta}^{(1)} + P_{\alpha\beta}^{(2)} +  P_{\alpha\beta}^{(3)} + \mathcal{O}(\lambda^4).
\end{equation}
The series expansions of the form given in equation~(\ref{series_expansion}) are carried out in terms of the small parameters $s_{13}$ and $\alpha$ which in turn are encoded in terms of the small parameter $\lambda$. This means that analytical probabilities presented in this work are driven by the oscillatory terms that depend on the large mass-squared difference $\Delta m_{31}^2$, whereas the small mass-squared difference $\Delta m_{21}^2$ affects the probability formulas through the parameter $\alpha$. The analytical probabilities are furthermore grouped by the available oscillatory frequencies and simplified until the shortest forms are obtained. The results are therefore superpositions of all different oscillation frequency modes. 

For convenience, we apply the following auxiliary parameters to maintain the results in a simple form
\begin{align}
    B &\equiv A-1-\gamma^2,\label{eq:B}\\
    C &\equiv A\gamma,\label{eq:C}\\
    D &\equiv (A-1)^2+\gamma^2.\label{eq:D}
\end{align}

The formulas that will be presented in this work are, to the best of our knowledge, the most compact and simplified forms for the neutrino oscillation probabilities.

\section{Oscillation probabilities with neutrino decay and matter effects}
\label{sec:expansions}

In this section, we present a complete set of series expansions for oscillation probabilities with neutrino decay. The probabilities are obtained with matter effects in section~\ref{sec:expansions:2}. The expansions are also presented in the vacuum limit $A \rightarrow 0$ and in unitarity relations, which are discussed in sections~\ref{sec:expansions:3} and~\ref{sec:expansions:4}, respectively.

\subsection{Neutrino oscillation and decay probabilities in matter}
\label{sec:expansions:2}
We first present the full set of neutrino oscillation probabilities in presence of neutrino decay and matter effects. The expansions are obtained in the scenario where the heaviest light neutrino mass state $\nu_3$ decays.

For $\nu_e \rightarrow \nu_e$ oscillations, we acquire the formula
\begin{equation}
    \label{Pee}
        P_{ee} = P_{ee}^{(0)} + P_{ee}^{(2)} = 1-\frac{2s_{13}^2}{D^2} \Bigg\{\Big(B^2-C^2+2ACD\Delta\Big)-\bigg[\big(B^2-C^2\big)\cos[2(A-1)\Delta]+2BC\sin\big[2(A-1)\Delta\big]\bigg]\text{e}^{-2\gamma\Delta}\Bigg\},
\end{equation}
where $P_{ee}^{(0)}$ and $P_{ee}^{(2)}$ and the zeroth-order and second-order corrections to the electron neutrino survival probability $P_{ee}$, respectively, whilst $\Delta \equiv \Delta m_{31}^2 L / (4 E)$ and $B$, $C$ and $D$ are functions of the decay parameter $\gamma$ and the matter potential $A$ as given in equations~(\ref{eq:B})--(\ref{eq:D}). The first-order term $P_{ee}^{(1)}$ does not exist in in equation~(\ref{Pee}). Here $P_{ee}^{(0)}$ equals to unity and $P_{ee}^{(2)}$ contains the terms that are proportional to $s_{13}^2$. The corrections are driven by the single oscillatory frequency $(A-1)\Delta$. As one can see, the expansion for $P_{ee}$ exists up to $\mathcal{O}(\lambda^2)$ and it is fully provided in equation~(\ref{Pee}). We note that the result is in agreement with that of Ref.~\cite{Chattopadhyay:2022ftv}, where the probability $P_{ee}$ was originally derived with the Cayley--Hamilton formalism. To the best of our knowledge, the expression presented in equation~(\ref{Pee}) is the shortest and the most compact expression available.

One of the notable features in the compact analytical formula~(\ref{Pee}) is the exponential factor $e^{-2\gamma\Delta}$, which appears in all of the oscillatory terms for $P_{ee}$. The introduction of neutrino decay gives rise to the exponential factor which in turn leads to the damping of the survival probability $P_{ee}$. As we will see in this work, the exponential factors are present for other neutrino oscillation channels as well.

In case of $\nu_e \rightarrow \nu_\mu$ oscillations, we obtain the compact formula
\begin{equation}
    \label{Pemu}
    \begin{split}
        P_{e\mu} &= P_{e\mu}^{(2)} + P_{e\mu}^{(3)} = s_{13}^2s_{23}^2\frac{1+\gamma^2}{D}\Big\{1-2\cos\big[2(A-1)\Delta\big]\text{e}^{-2\gamma\Delta}+\text{e}^{-4\gamma\Delta}\Big\}\\
        &+\frac{4\alpha s_{13}c_{12}s_{12}c_{23}s_{23}}{D}\frac{\sin(A\Delta)}{A} \Bigg\{B\bigg[\sin(A\Delta-\delta)+\sin\big[(A-2)\Delta+\delta\big]\text{e}^{-2\gamma\Delta}\bigg] \\
        &+C\bigg[\cos(A\Delta-\delta)-\cos\big[(A-2)\Delta+\delta\big]\text{e}^{-2\gamma\Delta}\bigg]\Bigg\},
    \end{split}    
\end{equation}
where the probability is provided up to $\mathcal{O}(\lambda^3)$. In this case, the zeroth-order and first-order corrections $P_{e\mu}^{(0)}$ and $P_{e\mu}^{(1)}$ are zero, whereas $P_{e\mu}^{(2)}$ and $P_{e\mu}^{(3)}$ are proportional to $s_{13}^2$ and $\alpha s_{13}$, respectively. Moreover, the second-order correction $P_{e\mu}^{(2)}$ contains only the oscillatory frequency $(A-1)\Delta$, while the third-order corrections $P_{e\mu}^{(3)}$ contain all three available frequencies. We moreover see exponentially damped factors in many of the oscillatory terms in equation~(\ref{Pemu}), but not in all of them. We find equation~(\ref{Pemu}) to be in agreement with the one derived in Ref.~\cite{Chattopadhyay:2022ftv}, where the expansion for $P_{e\mu}$ was originally derived using the Cayley--Hamilton formalism.

The compact analytical formula presented in equation~(\ref{Pemu}) shows the effect of the neutrino decay parameter $\gamma$ on the transition probability $P_{e\mu}$. This formula is obtained in the series expansion method assuming both $s_{13}$ and $\alpha$ to be small parameters. This method leads to the neutrino oscillation probability $P_{e\mu}$ to be driven by oscillatory frequencies such as $(A-1)\Delta$, which depends on the mass-squared difference $\Delta m_{31}^2$. The other mass-squared difference $\Delta m_{21}^2$ is also present through the parameter $\alpha$, which is assumed to be small but not zero. The analytical expression provided in equation~(\ref{Pemu}) therefore takes into account contributions from both $\Delta m_{31}^2$ and $\Delta m_{21}^2$. The same applies to all of the results that are presented in this work.

The series expansion for $\nu_e \rightarrow \nu_\tau$ oscillations up to $\mathcal{O}(\lambda^3)$ can be written as
\begin{align}
    \begin{split}
    \label{Petau}
        P_{e\tau} &= P_{e\tau}^{(2)} + P_{e\tau}^{(3)} = s_{13}^2c_{23}^2\frac{1+\gamma^2}{D}\Big\{1-2\cos\big[2(A-1)\Delta\big]\text{e}^{-2\gamma\Delta}+\text{e}^{-4\gamma\Delta}\Big\}\\
        &-\frac{4\alpha s_{13}c_{12}s_{12}c_{23}s_{23}}{D}\frac{\sin(A\Delta)}{A}\Bigg\{B\bigg[\sin(A\Delta-\delta)+\sin\big[(A-2)\Delta+\delta\big]\text{e}^{-2\gamma\Delta}\bigg] \\
        &+C\bigg[\cos(A\Delta-\delta)-\cos\big[(A-2)\Delta+\delta\big]\text{e}^{-2\gamma\Delta}\bigg]\Bigg\},
    \end{split}
\end{align}
where $P_{e\tau}^{(0)} = P_{e\tau}^{(1)} = 0$, $P_{e\tau}^{(2)} \propto s_{13}^2$ and $P_{e\tau}^{(3)} \propto \alpha s_{13}$. We note that the second-order correction $P_{e\tau}^{(2)}$ in equation~(\ref{Petau}) can be obtained by replacing $s_{23}^2$ with $c_{23}^2$ in the second-order correction $P_{e\mu}^{(2)}$ in equation~(\ref{Pemu}). The third-order correction $P_{e\tau}^{(3)}$ in equation~(\ref{Petau}) can furthermore be obtained from the third-order correction $P_{e\mu}^{(3)}$ in equation~(\ref{Pemu}) through the conversion $c_{23} s_{23} \rightarrow - s_{23} c_{23} = -c_{23} s_{23}$. The oscillatory frequencies are also the same in these two series expansions.

In the case of $\nu_\mu \rightarrow \nu_\mu$, we obtain the following expression
\begin{equation}
    \label{Pmumu0}
        P_{\mu\mu}^{(0)} = 1-s_{23}^2\Big(1-\text{e}^{-4\gamma\Delta}\Big) - c_{23}^2s_{23}^2\Big[1-2\cos(2\Delta)\text{e}^{-2\gamma\Delta}+\text{e}^{-4\gamma\Delta}\Big].
\end{equation}
The formula in equation~(\ref{Pmumu0}) shows the zeroth-order terms in the series expansion obtained for the muon neutrino survival probability $P_{\mu\mu}$. The probability has been previously presented in literature in many works, see {\em e.g.} Refs.~\cite{Barger:1998xk,Barger:1999bg,Lindner:2001fx}, and it has especially been confirmed to zeroth order in Ref.~\cite{Chattopadhyay:2022ftv}. The terms proportional to $s_{13}$ or $\alpha$ are not present in this equation, indicating $P_{\mu\mu}^{(0)} \propto \lambda^0$. We notice that $P_{\mu\mu}^{(0)}$ is driven by the oscillatory frequency $\Delta$.

We additionally present the higher-order terms for $P_{\mu\mu}$ up to $\mathcal{O}(\lambda^3)$. They are given by
\begin{equation}
    \label{Pmumu2}
    \begin{split}
        P_{\mu\mu}^{(2)} & = \frac{2s_{13}^2s_{23}^2}{D^2} \bigg\{\bigg[\left(B^2-C^2\right)\cos(2A\Delta)-2BC\sin(2A\Delta)\bigg]c_{23}^2 \\
        &+\bigg[\left(B^2-C^2\right)\cos\big[2(A-1)\Delta\big]-2BC\sin\big[2(A-1)\Delta\big]\bigg]s_{23}^2\text{e}^{-2\gamma\Delta}\\
        &-\bigg[\cos(2\Delta)c_{23}^2+s_{23}^2\text{e}^{-2\gamma\Delta}\bigg]\bigg(B^2-C^2-2ACD\Delta\bigg)\text{e}^{-2\gamma\Delta}\\
        &+2AB\big(\gamma+D\Delta\big)\sin(2\Delta)c_{23}^2\text{e}^{-2\gamma\Delta}\bigg\}+4\alpha c_{12}^2 c_{23}^2s_{23}^2\Delta\sin(2\Delta)\text{e}^{-2\gamma\Delta}
    \end{split}
\end{equation}
and
\begin{equation}
    \label{Pmumu3}
    \begin{split}
        P_{\mu\mu}^{(3)} &= \frac{4\alpha s_{13}c_{12}s_{12}c_{23}s_{23}\cos\delta}{AD} \Bigg\{A\big(1-A\big) s_{23}^2 \text{e}^{-4\gamma\Delta}+D\bigg[c_{23}^2-\cos(2\theta_{23})\cos(2\Delta)\text{e}^{-2\gamma\Delta}\bigg] \\
        &+ B\bigg[\cos(2A\Delta)c_{23}^2-\bigg(\cos(2\Delta)c_{23}^2-\cos\big[2(A-1)\Delta\big]s_{23}^2\bigg)\text{e}^{-2\gamma\Delta}\bigg] \\
        &- C\bigg[\sin(2A\Delta)c_{23}^2-\bigg(\sin(2\Delta)c_{23}^2+\sin\big[2(A-1)\Delta\big]s_{23}^2\bigg)\text{e}^{-2\gamma\Delta}\bigg] \Bigg\}.
    \end{split}
\end{equation}
The first-order term $P_{\mu\mu}^{(1)}$ is found to be zero and is therefore omitted. The higher-order terms shown in equations~(\ref{Pmumu2}) and (\ref{Pmumu3}) are presented for the first time in this work. One can immediately observe that the formula in equation~(\ref{Pmumu2}) consists of terms that are proportional to $s_{13}^2$, which therefore correspond to $\mathcal{O}(\lambda^2)$. Correspondingly, the third-order term shown in equation~(\ref{Pmumu3}) is proportional to $\alpha s_{13}$ and therefore matches $\mathcal{O}(\lambda^3)$. The oscillatory frequencies in equations~(\ref{Pmumu2}) and (\ref{Pmumu3}) are $\Delta, A\Delta$ and $(A-1)\Delta$. We find the probability formula presented in equations~(\ref{Pmumu0})--(\ref{Pmumu3}) for the survival probability $P_{\mu\mu}$ to be the shortest one available.

%It should be noticed that the analytical expressions presented in this section are grouped by oscillatory terms including $\Delta, A\Delta$ and $(A-1)\Delta$. One can always choose to arrange the terms in a different order, however, we find the forms presented in this work to be the shortest and most compact ones}.

For $\nu_\mu \rightarrow \nu_\tau$ oscillations, the zeroth-order term is
\begin{equation}
    \label{Pmutau0}
        P_{\mu\tau}^{(0)} = c_{23}^2s_{23}^2\Big[1-2\cos(2\Delta)\text{e}^{-2\gamma\Delta}+\text{e}^{-4\gamma\Delta}\Big].
\end{equation}
The first-order term is again zero and not shown here. To second-order, we reach
\begin{equation}
    \label{Pmutau2}
    \begin{split}
        P_{\mu\tau}^{(2)} &= -\frac{2s_{13}^2 c_{23}^2 s_{23}^2}{D^2} \bigg\{\bigg[\left(B^2-C^2\right)\cos(2A\Delta)-2BC\sin(2A\Delta) \bigg]\\
        &-\bigg[\left(B^2-C^2\right)\cos\big[2(A-1)\Delta\big]-2BC\sin\big[2(A-1)\Delta\big]\bigg]\text{e}^{-2\gamma\Delta} \\
        &-\bigg[\cos(2\Delta)-\text{e}^{-2\gamma\Delta}\bigg]\bigg(B^2-C^2-2ACD\Delta\bigg)\text{e}^{-2\gamma\Delta} \\ &+2AB\big(\gamma+D\Delta\big)\sin(2\Delta)\text{e}^{-2\gamma\Delta}\bigg\}-4\alpha c_{12}^2 c_{23}^2s_{23}^2\Delta\sin(2\Delta)\text{e}^{-2\gamma\Delta}.
    \end{split}
\end{equation}
The third-order term in $P_{\mu\tau}$ is more cumbersome. It can be written as
\begin{equation}
    \label{Pmutau3}
    \begin{split}
        P_{\mu\tau}^{(3)} &= \frac{2\alpha s_{13}c_{12}s_{12}c_{23}s_{23}}{AD} \bigg\{\bigg[A\big(1-A\big)\text{e}^{-4\gamma\Delta}-D\Big(1-2\cos(2\Delta)\text{e}^{-2\gamma\Delta}\Big)\bigg]\cos(2\theta_{23})\cos\delta-C\text{e}^{-4\gamma\Delta}\sin\delta \\
        &-B\bigg[\bigg(\cos(2A\Delta)-\Big[\cos(2\Delta)+\cos\big[2(A-1)\Delta\big]\Big]\text{e}^{-2\gamma\Delta}\bigg)\cos(2\theta_{23})\cos\delta\\ 
        &+\bigg(\sin(2A\Delta)-\Big[\sin(2\Delta) +\sin\big[2(A-1)\Delta\big]\Big]\text{e}^{-2\gamma\Delta}\bigg)\sin\delta\bigg]\\
        & +C \bigg[ \bigg(\sin(2A\Delta)-\Big[\sin(2\Delta)+\sin\big[2(A-1)\Delta\big]\Big]\text{e}^{-2\gamma\Delta}\bigg)\cos(2\theta_{23})\cos\delta \\
        &-\bigg(\cos(2A\Delta)-\Big[\cos(2\Delta)+\cos\big[2(A-1)\Delta\big]\Big]\text{e}^{-2\gamma\Delta}\bigg)\sin\delta \bigg] \bigg\}.
    \end{split}
\end{equation}
Equations~(\ref{Pmutau0})--(\ref{Pmutau3}) show the probability for $\nu_\mu \rightarrow \nu_\tau$ transition up to $\mathcal{O}(\lambda^3)$. The expansions are novel results and they have not been presented elsewhere. The zeroth-order term in equation~(\ref{Pmutau0}) is independent of $s_{13}$ and $\alpha$ and corresponds to $\mathcal{O}(\lambda^0) = \mathcal{O}(1)$. The second-order term in equation~(\ref{Pmutau2}) is proportional to $s_{13}^2$ and therefore corresponds to $\mathcal{O}(\lambda^2)$. Similarly, the third-order term in equation~(\ref{Pmutau3}) is driven by $\alpha s_{13}$ and matches $\mathcal{O}(\lambda^3)$. Thus, the sole oscillatory frequency in the zeroth-order term $P_{\mu\tau}^{(0)}$ is again $\Delta$, whereas the oscillatory frequencies in the second-order and third-order terms $P_{\mu\tau}^{(2)}$ and $P_{\mu\tau}^{(3)}$ are $\Delta, A\Delta$ and $(A-1)\Delta$.

We finally remark that the corresponding formulas for the antineutrino oscillation probabilities $P_{\bar{\alpha}\bar{\beta}}$, where $\bar{\alpha}$ and $\bar{\beta}$ indicate antineutrino flavors, can be obtained from the neutrino oscillation probabilities $P_{\alpha \beta}$ in equations~(\ref{Pee})--(\ref{Pmutau3}) through the transformations $\delta \rightarrow -\delta$ and $A \rightarrow -A$, {\em i.e.} $P_{\bar{\alpha}\bar{\beta}} = P_{\alpha \beta}(\delta \rightarrow -\delta, A \rightarrow -A)$. The auxiliary parameters for those formulas are given by \begin{align}
\bar{B} &= -A - 1 - \gamma^2,\\
\bar{C} &= -A\gamma,\\
\bar{D} &= (A+1)^2 + \gamma^2.
\end{align}

\subsection{Vacuum limits of the neutrino oscillation and decay probabilities}
\label{sec:expansions:3}
We now obtain the vacuum limits for the neutrino oscillation probabilities with invisible neutrino decay. The vacuum limits are acquired by applying $A \rightarrow 0$ in the series expansions presented in equations~(\ref{Pee})--(\ref{Pmutau3}).

For the $\nu_e \rightarrow \nu_e$ case, the vacuum probability reads as
\begin{equation}
    \label{vacPee}
        P_{ee}^\text{vac} = 1-2s_{13}^2\Big[1-\cos(2\Delta)\text{e}^{-2\gamma\Delta}\Big].
\end{equation}
The probabilities for $\nu_e \rightarrow \nu_\mu$ and $\nu_e \rightarrow \nu_\tau$ channels are similarly given by the following equations
\begin{equation}
    \label{vacPemu}
        P_{e\mu}^\text{vac} = s_{13}^2 s_{23}^2\Big[1-2\cos(2\Delta)\text{e}^{-2\gamma\Delta}+\text{e}^{-4\gamma\Delta}\Big]+4\alpha s_{13}c_{12}s_{12}c_{23}s_{23} \Delta\Big[\sin(2\Delta-\delta)+\text{e}^{-2\gamma\Delta}\sin\delta\Big],
\end{equation}
\begin{equation}
    \label{vacPetau}
        P_{e\tau}^\text{vac} = s_{13}^2c_{23}^2\Big[1-2\cos(2\Delta)\text{e}^{-2\gamma\Delta}+\text{e}^{-4\gamma\Delta}\Big]-4\alpha s_{13}c_{12}s_{12}c_{23}s_{23}\Delta\Big[\sin(2\Delta-\delta)+\text{e}^{-2\gamma\Delta}\sin\delta\Big].
\end{equation}
For $\nu_\mu \rightarrow \nu_\mu$ and $\nu_\mu \rightarrow \nu_\tau$ channels, we arrive at
\begin{equation}
    \label{vacPmumu}
    \begin{split}
        P_{\mu\mu}^\text{vac} &=  1-s_{23}^2\Big(1-\text{e}^{-4\gamma\Delta}\Big)-c_{23}^2 s_{23}^2\Big[1-2\cos(2\Delta)\text{e}^{-2\gamma\Delta}+\text{e}^{-4\gamma\Delta}\Big]+4\alpha c_{12}^2c_{23}^2s_{23}^2 \Delta\sin(2\Delta)\text{e}^{-2\gamma\Delta} \\
        &+s_{13}^2 s_{23}^2 \Big\{1-\text{e}^{-4\gamma\Delta}+\cos(2\theta_{23})\Big[1-2\cos(2\Delta)\text{e}^{-2\gamma\Delta}+\text{e}^{-4\gamma\Delta}\Big]\Big\} \\
        &-\frac{2\alpha s_{13} c_{12}s_{12}c_{23}s_{23}\cos\delta}{(1+\gamma^2)}\bigg\{1-\text{e}^{-4\gamma\Delta}+\cos(2\theta_{23})\Big[1-2\cos(2\Delta)\text{e}^{-2\gamma\Delta}+\text{e}^{-4\gamma\Delta}\Big]\\
        &-2\sin(2\Delta)\text{e}^{-2\gamma\Delta}\Big[\cos(2\theta_{23})\Big(\big(1+\gamma^2\big)\Delta+\gamma\Big)-1\Big]\bigg\},
    \end{split}
\end{equation}
\begin{equation}
    \label{vacPmutau}
    \begin{split}
        P_{\mu\tau}^\text{vac} &= c_{23}^2s_{23}^2\Big[1-2\cos(2\Delta)\text{e}^{-2\gamma\Delta}+\text{e}^{-4\gamma\Delta}\Big]-4\alpha c_{12}^2c_{23}^2s_{23}^2 \Delta\sin(2\Delta)\text{e}^{-2\gamma\Delta}\\ 
        &-2s_{13}^2c_{23}^2 s_{23}^2 \Big[1-2\cos(2\Delta)\text{e}^{-2\gamma\Delta}+\text{e}^{-4\gamma\Delta}\Big]\\
        &+\frac{2\alpha s_{13} c_{12}s_{12}c_{23}s_{23}}{(1+\gamma^2)}\bigg\{\sin\delta\Big(1-\text{e}^{-2\gamma\Delta}\Big)\Big[2\Delta(1+\gamma^2)-\gamma\Big(1-\text{e}^{-2\gamma\Delta}\Big)\Big]\\
        &-\cos(2\theta_{23})\cos\delta\Big[\Big(1-2\cos(2\Delta)\text{e}^{-2\gamma\Delta}+\text{e}^{-4\gamma\Delta}\Big)+2\big(1+\gamma^2\big)\Delta\sin(2\Delta)\text{e}^{-2\gamma\Delta}\Big]\bigg\}.        
    \end{split}
\end{equation}

%We finally present the vacuum limit for the $\nu_\tau \rightarrow \nu_\tau$ channel
%\begin{equation}
%    \label{vacPtautau}
%    \begin{split}
%        P_{\tau\tau}^\text{vac} &= 1-c_{23}^2\Big(1-\text{e}^{-4\gamma\Delta}\Big)-c_{23}^2s_{23}^2\Big[1-2\cos(2\Delta)\text{e}^{-2\gamma\Delta}+\text{e}^{-4\gamma\Delta}\Big]+4\alpha c_{12}^2c_{23}^2s_{23}^2 \Delta\sin(2\Delta)\text{e}^{-2\gamma\Delta}\\
%        &+s_{13}^2 c_{23}^2 \Big\{1-\text{e}^{-4\gamma\Delta}+\cos(2\theta_{23})\Big[1-2\cos(2\Delta)\text{e}^{-2\gamma\Delta}+\text{e}^{-4\gamma\Delta}\Big]\Big\} \\
%        &+\frac{2\alpha s_{13}  c_{12}s_{12}c_{23}s_{23} \cos\delta}{(1+\gamma^2)}\bigg\{1-\text{e}^{-4\gamma\Delta}+\cos(2\theta_{23})\Big[1-2\cos(2\Delta)\text{e}^{-2\gamma\Delta}+\text{e}^{-4\gamma\Delta}\Big]\\&+2\sin(2\Delta)\text{e}^{-2\gamma\Delta}\Big[2c_{23}^2\big(1+\gamma^2\big)\Delta-\gamma\Big]\bigg\}.
%    \end{split}
%\end{equation}

\subsection{Unitarity relations with neutrino decay}
\label{sec:expansions:4}

Including invisible decay into neutrino oscillation probabilities lead to violation of unitarity. In this section, we derive the unitarity relations for the $e$, $\mu$ and $\tau$ channels. The results are presented in equations~(\ref{uniPea})--(\ref{uniPtaua}):
\begin{equation}
    \label{uniPea}
        \sum_\beta P_{e \beta} = 1+\frac{s_{13}^2}{D^2}\bigg\{3C^2-B^2-4ACD\Delta+(1+\gamma^2)D\text{e}^{-4\gamma\Delta}+4C\text{e}^{-2\gamma\Delta}\Big[B\sin\big[2(A-1)\Delta]-C\cos\big[2(A-1)\Delta]\Big]\bigg\},
\end{equation}

\begin{equation}
    \label{uniPmua}
    \begin{split}
        \sum_\beta P_{\mu \beta} &= 1-\big(1-\text{e}^{-4\gamma\Delta}\big)s_{23}^2-\frac{s_{13}^2s_{23}^2}{D^2}\bigg\{\big(1-\text{e}^{-4\gamma\Delta}\big)(1+\gamma^2)(A+B)-A^2\Big[(1+\gamma^2)-(1-3\gamma^2)\text{e}^{-4\gamma\Delta}\Big]\\
        &-4ACD\Delta \text{e}^{-4\gamma\Delta}+4C\text{e}^{-2\gamma\Delta}\Big[B\sin\big[2(A-1)\Delta\big]+C\cos\big[2(A-1)\Delta\big] \Big]\bigg\}\\
        &-\frac{2\alpha s_{13} c_{12}s_{12}c_{23}s_{23}}{AD}\bigg\{C\Big[1-2\cos\big[2(A-1)\Delta\big]\text{e}^{-2\gamma\Delta}+\text{e}^{-4\gamma\Delta}\Big]\sin\delta\\
        &-\Big[A(A-1)\big(1-\text{e}^{-4\gamma\Delta}\big)-2C\cos(2\theta_{23})\sin\big[2(A-1)\Delta\big]\text{e}^{-2\gamma\Delta}\Big]\cos\delta\bigg\}
    \end{split}
\end{equation}
and
\begin{equation}
    \label{uniPtaua}
    \begin{split}
        \sum_\beta P_{\tau \beta} &= 1-\big(1-\text{e}^{-4\gamma\Delta}\big)c_{23}^2-\frac{s_{13}^2c_{23}^2}{D^2}\bigg\{\big(1-\text{e}^{-4\gamma\Delta}\big)(1+\gamma^2)(A+B)-A^2\Big[(1+\gamma^2)-(1-3\gamma^2)\text{e}^{-4\gamma\Delta}\Big]\\
        &-4ACD\Delta \text{e}^{-4\gamma\Delta}+4C\text{e}^{-2\gamma\Delta}\Big[B\sin\big[2(A-1)\Delta\big]+C\cos\big[2(A-1)\Delta\big]\Big]\bigg\}\\
        &+\frac{2\alpha s_{13} c_{12}s_{12}c_{23}s_{23}}{AD}\bigg\{C\Big[1-2\cos\big[2(A-1)\Delta\big]\text{e}^{-2\gamma\Delta}+\text{e}^{-4\gamma\Delta}\Big]\sin\delta\\
        &-\Big[A(A-1)\big(1-\text{e}^{-4\gamma\Delta}\big)+2C\cos(2\theta_{23})\sin\big[2(A-1)\Delta\big]\text{e}^{-2\gamma\Delta} \Big]\cos\delta\bigg\}.   
    \end{split}
\end{equation}
We observe that the transposed unitarity relations $\sum_\alpha P_{\alpha e}$,  $\sum_\alpha P_{\alpha \mu}$ and $\sum_\alpha P_{\alpha \tau}$, where the two indices each neutrino oscillation probability are interchanged in comparison to the original one, can be obtained by applying the transformation $\delta \rightarrow -\delta$ in equations~(\ref{uniPea})--(\ref{uniPtaua}), respectively. We can immediately see that the unitarity relations reduce to unity in both cases when $\gamma \rightarrow 0$.

\section{Discussion}
\label{Discussion}

In this section, we discuss the qualitative effects that occur for the neutrino oscillation probabilities in presence of neutrino decay. We begin by obtaining the probability formulas presented in section~\ref{sec:expansions} at the no-decay limit and comparing the results with those derived in Ref.~\cite{Akhmedov:2004ny}. We also investigate the violation of unitarity and estimate the accuracy of the expansion formulas presented in this work.

\subsection{No-decay limit}
\label{Limits}

At the no-decay limit $\gamma\rightarrow 0$, the analytical probabilities presented in this work are expected to follow the standard three-flavor paradigm. The auxiliary parameters $B$, $C$ and $D$ introduced in equations~(\ref{eq:B})--(\ref{eq:D}) then read as
\begin{align}
        B_{\gamma\to 0} &= A-1,\\
        C_{\gamma\to 0} &= 0,\\
        D_{\gamma\to 0} &= \left(A-1\right)^2.
\end{align}
One can readily see that $B$, $C$ and $D$ are finite at the no-decay limit. The vanishing parameter $C$ is furthermore never present in denominators and does not induce divergences in the probability formulas.

We take a moment to compare our results in the no-decay limit with the analytical probabilities that were derived in Ref.~\cite{Akhmedov:2004ny}. As the authors of Ref.~\cite{Akhmedov:2004ny} used the same methodology, that is, obtaining analytical expressions for the neutrino oscillation probabilities via perturbative diagonalization in Cayley--Hamilton formalism, it is expected that the results derived in our work agree with those presented in Ref.~\cite{Akhmedov:2004ny}. Indeed, we find that the expansions that we presented for the probabilities $P_{ee}$, $P_{e\mu}$ and $P_{e\tau}$ in equations~(\ref{Pee}), (\ref{Pemu}) and (\ref{Petau}), respectively, are consistent with the probability formulas discussed in Ref.~\cite{Akhmedov:2004ny}. Our findings for the probabilities $P_{\mu\mu}$ and $P_{\mu\tau}$ in equations~(\ref{Pmumu0})--(\ref{Pmumu3}) and (\ref{Pmutau0})--(\ref{Pmutau3}) on the other hand are not in perfect agreement with the expansions presented in Ref.~\cite{Akhmedov:2004ny}. When the results for $P_{\mu\mu}$ and $P_{\mu\tau}$ are compared to the results in Ref.~\cite{Akhmedov:2004ny}, we find that other third-order terms are present in the two probabilities. In addition, this finding applies to $P_{\tau\tau}$, which can be directly computed as $P_{\tau\tau} = P_{\mu\mu}(s_{23}^2 \leftrightarrow c_{23}^2, \sin2\theta_{23} \rightarrow -\sin2\theta_{23})$~\cite{Akhmedov:2004ny}. The difference between the two sets of probabilities can be exemplified and characterized by the following contribution
\begin{equation}
    \label{error}
        \Delta P_{\alpha \beta}^\text{no decay} = \alpha s_{13} \sin(2\theta_{12})\sin(4\theta_{23})\sin^2\Delta \cos\delta
        = 8\alpha s_{13}c_{12}s_{12}c_{23}s_{23}\cos(2\theta_{23})\sin^2\Delta\cos\delta,
\end{equation}
where $\alpha, \beta = \mu, \tau$. This additional contribution (or term in the probability difference) is not present in Ref.~\cite{Akhmedov:2004ny} and it originates from expanding to lower order in $\alpha$ while applying perturbative diagonalization to the effective Hamiltonian. When we take $\alpha \sim \mathcal{O}(\lambda^2)$, the terms involving $\alpha^2$ ($\sim\mathcal{O}(\lambda^4$)) are disregarded in the energy eigenvalues~(\ref{energyeigenvalues1})--(\ref{energyeigenvalues3}). The first energy eigenvalue for example reads\footnote{Note that neutrino oscillation probabilities are finite at the no-decay limit $\gamma \rightarrow 0$ even when $A \rightarrow 0$ or $A \rightarrow 1$~\cite{Akhmedov:2004ny}.}
\begin{equation}
    \label{eq:discussion:2}
    E_1 = \frac{\Delta m_{31}^2}{2E}\bigg[A+\alpha s_{12}^2+s_{13}^2\frac{A(1-i \gamma)}{A-1+i \gamma}\bigg],%+\mathcal{O}(\lambda^4),
\end{equation}
whereas the corresponding one in Ref.~\cite{Akhmedov:2004ny} is given by
\begin{equation}
    \label{eq:discussion:3}
    E_1 \simeq \frac{\Delta m_{31}^2}{2E}\left[A+\alpha s_{12}^2+s_{13}^2\frac{A}{A-1}+\alpha^2\frac{\sin^2(2\theta_{12})}{4A}\right].
\end{equation}
The fact that the last term proportional to $\alpha^2$ is present in equation~(\ref{eq:discussion:3}), but not in equation~(\ref{eq:discussion:2}), is not surprising, since equation~(\ref{eq:discussion:3}) was derived up to second order in both small parameters $s_{13}$ ($\sim \mathcal{O}(\lambda)$) and $\alpha$ ($\sim \mathcal{O}(\lambda^2)$) separately \cite{Akhmedov:2004ny}. It is a trade-off that results from simplifying the complicated expressions that arise from calculating neutrino oscillation probabilities with neutrino decay. By expanding the exponentials in the time-evolution operator as functions of $\alpha^2$ in energy eigenvalues $E_1$, $E_2$ and $E_3$, we obtain additional terms that are linear in $\alpha$ from the chain rule. In some instances these extra terms multiply by $s_{13}$ to generate third-order terms in the small parameter $\lambda$ such as a term proportional to $\alpha s_{13}$ ($\sim \mathcal{O}(\lambda^3)$). This was not taken into account in the derivation of the probability formulas in the present work, but we consider it the most probable source to the contribution with the mixed factor $\alpha s_{13}$ that is presented in equation~(\ref{error}). Moreover, this is also likely the case in Ref.~\cite{Chattopadhyay:2022ftv}, where authors reached the same energy eigenvalues as those we presented in equations~(\ref{energyeigenvalues1})--(\ref{energyeigenvalues3}). We estimate the magnitude of equation~(\ref{error}) to be $10^{-3}$ using the experimentally determined values for the neutrino oscillation parameters~\cite{deSalas:2020pgw}.

Some expansions of neutrino oscillation probabilities with neutrino decay have been performed in vacuum. An example of such publication is Ref.~\cite{Choubey:2020dhw}, where authors present vacuum formulas for $P_{\mu e}$ and $P_{\mu\mu}$ using the so-called one-mass-squared-difference (OMSD) approximation. In this scenario, expansions are carried out assuming $\Delta m_{21}^2=0$. We find that our results for $P_{\mu e}$ and $P_{\mu\mu}$ are in agreement with those derived in Ref.~\cite{Choubey:2020dhw} in the vacuum limit $A \rightarrow 0$.

\subsection{Violation of unitarity}
\label{Violation of unitarity}
In the absence of neutrino decay, the oscillation probabilities for neutrinos satisfy the unitarity relations
\begin{equation}
\label{Unitarity relations}
    \sum_{\alpha}P_{\alpha\beta} = \sum_\beta P_{\alpha\beta} = 1,
\end{equation}
where $\alpha$ and $\beta$ run over the flavor states $e, \mu$ and $\tau$. This ensures that each neutrino must either remain in its original state or undergo a transition to another flavor state.

In perturbative expansion in small parameters, the unitarity relations shown in equation~(\ref{Unitarity relations}) are satisfied up to a given order. It has been shown that the number of independent probabilities can be reduced to five~\cite{Akhmedov:2004ny}. When neutrino decay is allowed, however, the evolution is no longer unitary. Equations~(\ref{unitaritytable1})--(\ref{unitaritytable3}) show the corresponding unitarity relations when probabilities derived under neutrino decay are included: 
\begin{align}
    \sum_\beta P_{e\beta} &= 1+\mathcal{O}(\lambda^2),\label{unitaritytable1}\\
    \sum_\beta P_{\mu \beta} &= 1-\big(1-\text{e}^{-4\gamma\Delta}\big)s_{23}^2+\mathcal{O}(\lambda^2),\label{unitaritytable2}\\
    \sum_\beta P_{\tau \beta} &= 1-\big(1-\text{e}^{-4\gamma\Delta}\big)c_{23}^2+\mathcal{O}(\lambda^2).\label{unitaritytable3}
    %\label{Unitarity table}
\end{align}
It is observed that all three equations display corrections of order $\mathcal{O}(\lambda^2)$. The sums in the second and third equations~(\ref{unitaritytable2}) and~(\ref{unitaritytable3}) additionally contain the zeroth-order corrections $(1-\text{e}^{-4\gamma\Delta})s_{23}^2$ and $(1-\text{e}^{-4\gamma\Delta})c_{23}^2$, respectively.

In the Cayley--Hamilton formalism, the reversed transition probabilities are obtained by simply replacing $\delta \rightarrow -\delta$, whereby $P_{\alpha\beta}(-\delta)=P_{\beta\alpha}(\delta)$. This relationship prevails in spite of unitarity violation. As the survival probability formulas $P_{ee}$ and $P_{\mu\mu}$ derived in this work and also $P_{\tau\tau}$ only depend on $\delta$ through $\cos\delta$, they are unchanged under the transformation $\delta \rightarrow -\delta$. The relations corresponding to the sums $\sum_\alpha P_{\alpha \beta}$ are therefore omitted.

We calculate the neutrino oscillation probabilities using the probability formulas~(\ref{Pee})--(\ref{Pmutau3}) in section~\ref{sec:expansions}. {We moreover derive the series expansion for $P_{\tau\tau}$ to be able to present it. We use the standard oscillation parameters $\theta_{12}$, $\theta_{13}$, $\theta_{23}$, $\delta$, $\Delta m_{21}^2$ and $\Delta m_{31}^2$ and assume the values that are listed in Table~\ref{Parameter table}. These values are adopted from Ref.~\cite{deSalas:2020pgw} assuming normal neutrino mass ordering. The allowed parameter values are also shown at 1$\sigma$, 2$\sigma$ and 3$\sigma$ C.L. For the decay parameter $\gamma$, we choose the value $0.1$. This value was found to be compatible with the atmospheric neutrino oscillation data from Super-Kamiokande~\cite{Gonzalez-Garcia:2008mgl}. The unitarity relations resulting from this setup are presented in Figure~\ref{fig:unitarity}.

As we can already tell from equations~(\ref{unitaritytable1})--(\ref{unitaritytable3}), the first equation has corrections of second order and higher. This indicates that unitarity is only minutely violated and the probability sum $P_{ee} + P_{e\mu} + P_{e\tau}$ is close to unity. This behavior can also be confirmed in Figure~\ref{fig:unitarity}. We note that the deviation from unitarity is the largest for the lowest energies, where the sum increases monotonically. In equations~(\ref{unitaritytable2}) and~(\ref{unitaritytable3}), we observe a more significant departure from unitarity at low energies. The comparatively higher impact on $P_{\mu e} + P_{\mu \mu} + P_{\mu \tau}$ and $P_{\tau e} + P_{\tau \mu} + P_{\tau \tau}$ is also evident in equations~(\ref{unitaritytable2}) and~(\ref{unitaritytable3}). In this regard, the zeroth-order corrections proportional to $1-\text{e}^{-4\gamma\Delta}$ become notable at low energies where $\Delta$ is large. Moreover, it is observed that violation of unitarity is most profound in equation~(\ref{unitaritytable2}), where muon neutrino survival and conversion probabilities are summed over. The qualitative behavior is similar for the probabilities in equation~(\ref{unitaritytable3}). Examining the full expressions in section \ref{sec:expansions:4}, we note that these probabilities are indeed nearly identical. Since we have assumed $\theta_{23}$ to be 49$^\circ$, we have $\cos\theta_{23}<\sin\theta_{23}$ and equation~(\ref{unitaritytable3}) becomes larger than equation~(\ref{unitaritytable2}). We also see this behavior in Figure~\ref{fig:unitarity}. We finally remark that the results presented in this work were obtained for constant matter density. The numerical results presented here do not take into account the uncertainties that may arise from the constant matter density assumption.

\begin{table}[!t]
    \centering
    \def\arraystretch{1.4}
    \begin{tabular}{c c c c c}
        \hline
        \textbf{Parameter} & \textbf{Unit} & \textbf{Best fit $\bm{\pm 1\sigma}$} & $\bm{2\sigma}$ \textbf{range} & $\bm{3\sigma}$ \textbf{range}  \\ 
        \hline 
        $\Delta m_{21}^2$ & $10^{-5}\text{eV}^2$ & $7.50_{-0.20}^{+0.22}$ & $7.12-7.93$ & $6.94-8.14$ \\
        $|\Delta m_{31}^2|$ & $10^{-3}\text{eV}^2$ & $2.55_{-0.03}^{+0.02}$ & $2.49-2.60$ & $2.47-2.63$ \\
        $\theta_{12}$ & $^\circ$ & $34.3\pm1.0$ & $32.3-36.4$ & $31.4-37.4$ \\
        $\theta_{13}$ & $^\circ$ & $8.53_{-0.12}^{+0.13}$ & $8.27-8.79$ & $8.13-8.92$ \\
        $\theta_{23}$ & $^\circ$ & $49.26\pm0.79$ & $47.37-50.71$ & $41.20-51.33$ \\
        $\delta$ & $\pi$ & $1.08_{-0.12}^{+0.13}$ & $0.84-1.42$ & $0.71-1.99$ \\ \hline
    \end{tabular}
    \caption{The values of the neutrino oscillation parameters used in this work. This table is adopted from Ref.~\cite{deSalas:2020pgw}.}
    \label{Parameter table}
\end{table}

\begin{figure}[!t]
    \centering
    \includegraphics[width=0.75\textwidth]{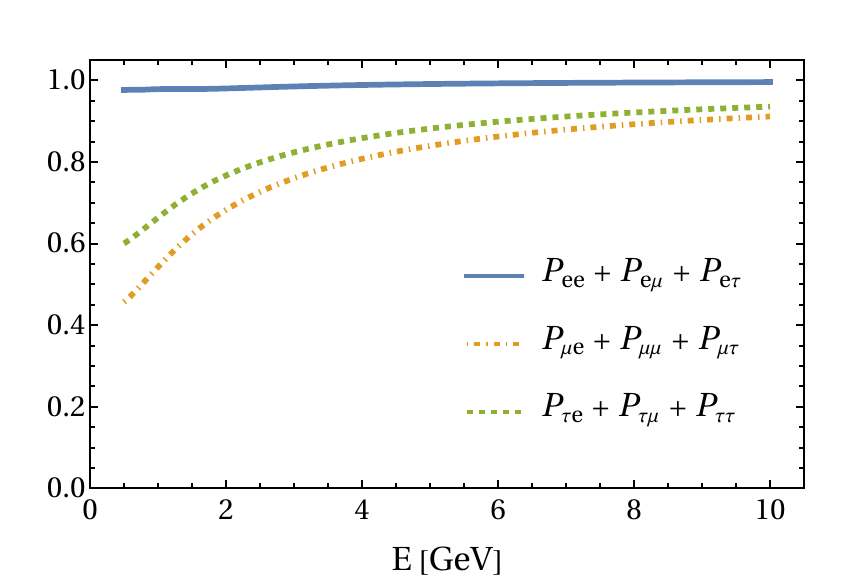}
    \caption{Unitarity relations $\sum_\beta P_{\alpha\beta}$ plotted in the energy range $0.5$~GeV--$10$~GeV and baseline length $1300$~km, where $\alpha, \beta = e, \mu, \tau$. The neutrino oscillation probabilities are computed for decay parameter $\gamma = 0.1$ and matter density $\rho = 3$~g/cm$^3$.}
    \label{fig:unitarity}
\end{figure}

\subsection{Accuracy of the series expansions}
\label{Accuracy sec}
We now investigate the accuracy of the series expansions that we presented in section~\ref{sec:expansions}. The accuracy of the series expansion method is discussed in great detail in Ref.~\cite{Akhmedov:2004ny}. In the present work, we use the double expansion method where both $s_{13}$ and $\alpha$ are used as expansion parameters. It is argued in Ref.~\cite{Akhmedov:2004ny} that this method is more accurate than the ones where either $s_{13}$ or $\alpha$ is considered as the expansion parameter. The expansions involving both $s_{13}$ and $\alpha$ are valid when
\begin{equation}
    \label{Accuracy condition}
    \alpha\Delta=\frac{
\Delta m_{21}^2L}{4E} \ll 1 \quad \iff \quad \frac{L}{E} \ll 10^{4} \text{ km/GeV}.
\end{equation}
Conversely, the probability formulas derived in this work become inaccurate when either the baseline length $L$ is very long or the neutrino energy $E$ is very low. 

Figure~\ref{fig:Probabilities} illustrates the accuracy of the probabilities that are obtained from the analytical probabilities derived in this work. The figure shows the approximate probabilities for decay parameter $\gamma = 0.1$, baseline length $L = 1300$~km and neutrino energies $E = 0.5$~GeV--$10$~GeV. The configuration corresponds to that of the DUNE setup. The probabilities are calculated from equations~(\ref{Pee})--(\ref{Pmutau3}) and they are depicted with blue solid lines. The figure also shows probabilities that have been obtained numerically from the time-evolution operator as described in section~\ref{sec:background}. The numerically computed probabilities are shown with yellow dashed lines. We finally show the numerically obtained neutrino oscillation probabilities for the standard case where there is no neutrino decay. The no-decay probabilities are presented with black solid lines.

% PROBABILITY PLOTS
%\textbf{Probabilities}
\begin{figure}[!t]
  \centering
  \begin{subfigure}{.48\linewidth}
    \centering
    \includegraphics[width = 1.\linewidth]{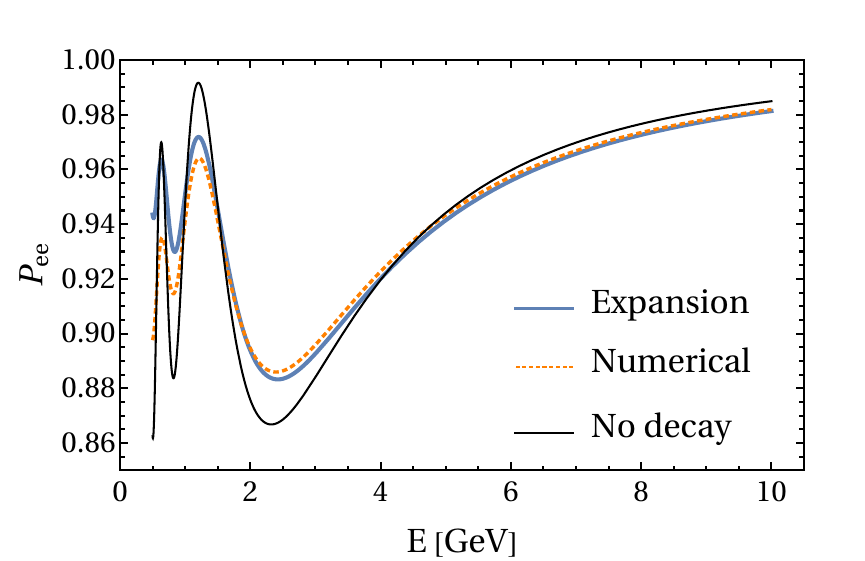}
    \caption{\label{fig:2a}Survival probability $P_{ee}$.}
  \end{subfigure}%
  \hspace{0.8em}% Space between image A and B
  \begin{subfigure}{.48\linewidth}
    \centering
    \includegraphics[width = 1.\linewidth]{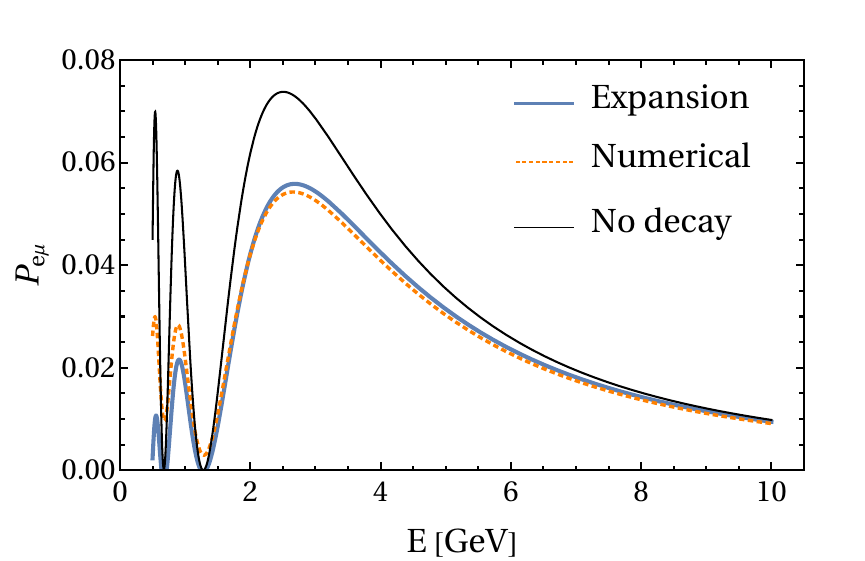}
    \caption{\label{fig:2b}Transition probability $P_{e\mu}$.}
  \end{subfigure}\\
    \begin{subfigure}{.48\linewidth}
    \centering
    \includegraphics[width = 1.\linewidth]{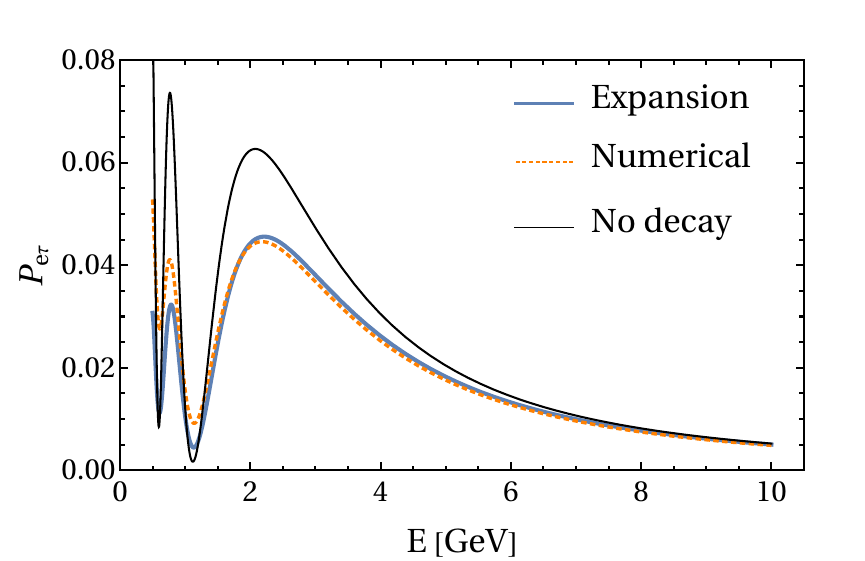}
    \caption{\label{fig:2c}Transition probability $P_{e\tau}$.}
  \end{subfigure}
    \hspace{0.8em}% Space between image A and B
    \begin{subfigure}{.48\linewidth}
    \centering
    \includegraphics[width = 1.\linewidth]{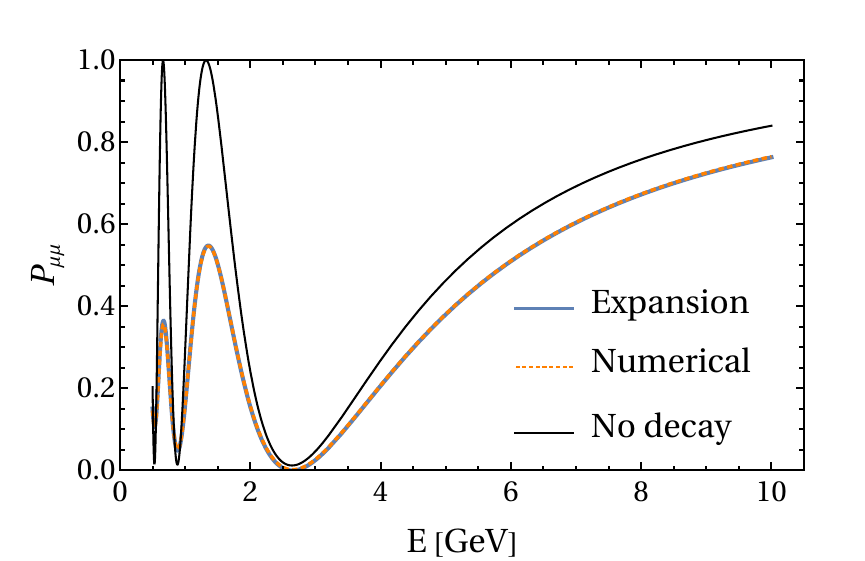}
    \caption{\label{fig:2d}Survival probability $P_{\mu\mu}$.}
  \end{subfigure}\\
      \begin{subfigure}{.48\linewidth}
    \centering
    \includegraphics[width = 1.\linewidth]{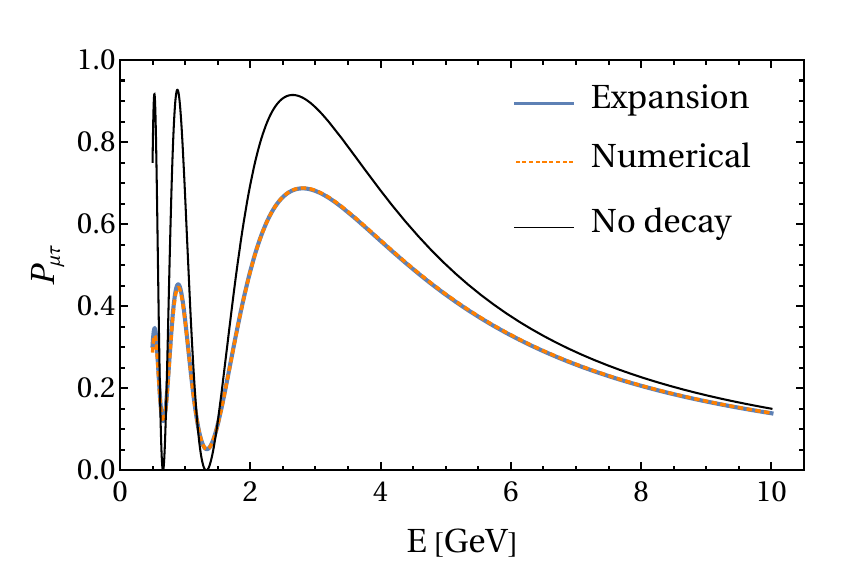}
    \caption{\label{fig:2e}Transition probability $P_{\mu\tau}$.}
  \end{subfigure}
    \hspace{0.8em}% Space between image A and B
    \begin{subfigure}{.48\linewidth}
    \centering
    \includegraphics[width = 1.\linewidth]{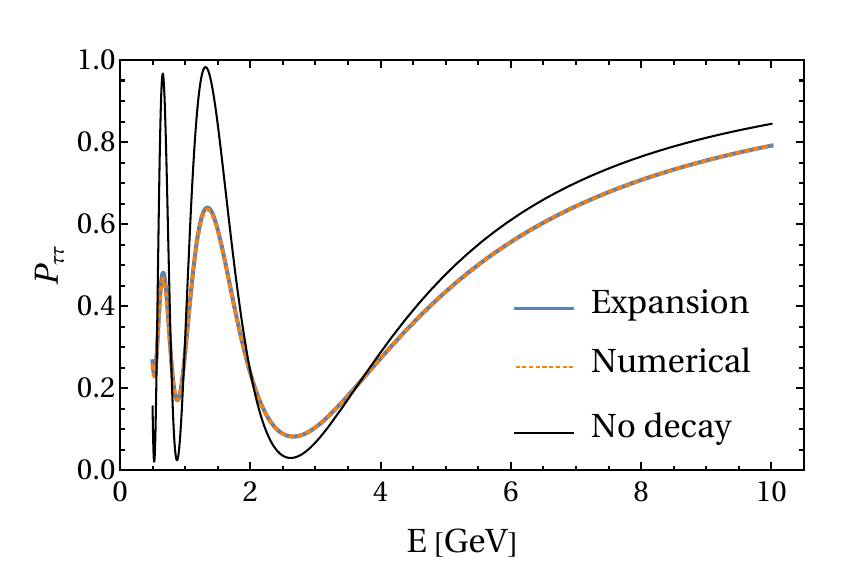}
    \caption{\label{fig:2f}Survival probability $P_{\tau\tau}$.}
  \end{subfigure}\\
  \caption{Neutrino oscillation probabilities with invisible decay neutrino decay of $\gamma =$ 0.1. The probabilities have been plotted for the series expansions (blue solid line) for neutrino energies $E = 0.5$~GeV--$10$~GeV and baseline length $L=1300$~km. For comparison, numerically computed probabilities are also plotted for neutrino decay (yellow dashed line) and for no decay (black solid line).}
  \label{fig:Probabilities}
\end{figure}

The accuracy condition shown in equation~(\ref{Accuracy condition}) is in agreement with the probabilities shown in Figure~\ref{fig:Probabilities}. The numerically computed probabilities are notably different from the approximate ones only at low energies where $E \lesssim 1.5$~GeV. We note that the discrepancy appears more profound in the electron neutrino probabilities $P_{ee}$, $P_{e\mu}$ and $P_{e\tau}$, where the probabilities are constrained to a narrow interval. In general, we find the analytically computed probabilities to be in a good agreement with the numerically calculated ones in the DUNE setup. It is also noteworthy that the neutrino oscillation probabilities yielded by the series expansions are closer to the numerical results obtained for neutrino decay than the numerical results obtained for no decay.

We notice that the survival probability $P_{ee}$ exhibits significant damping in Figure~\ref{fig:2a}, where changes in $P_{ee}$ are less steep than the corresponding probabilities obtained without neutrino decay. The damping effect originates from the oscillatory terms having the exponential factor $e^{-2\gamma\Delta}$ in equation~(\ref{Pee}). When the decay parameter $\gamma$ is non-zero, the damping suppresses the changes in $P_{ee}$, which lowers the probability near the oscillation maxima and increasing it near the oscillation minima. This behaviour can be observed in Figure~\ref{fig:2a}.

To illustrate the importance of the different terms appearing in the analytical formulas, we have plotted the three transition probabilities $P_{e\mu}$, $P_{e\tau}$ and $P_{\mu\tau}$ in different orders of $\lambda$ in Figure~\ref{fig:Corrections}. Each panel in this figure shows the contributions that arise from the zeroth-order term (blue solid lines) as well as from the terms up to the second order (yellow dashed lines) and up to the third order (magenta dot-dashed lines). The numerically calculated probabilities (black solid lines) are also presented in these three panels. The zeroth-order terms exist only for the transition probability $P_{\mu\tau}$, while the second-order and third-order corrections are present for all three transition probabilities. Figures~\ref{fig:Pemu-corr} and \ref{fig:Petau-corr} for $P_{e\mu}$ and $P_{e\tau}$, respectively, indicate that the probabilities which are plotted up to the third order in $\lambda$ approximate the numerically computed probabilities better than the probabilities plotted only up to the second order. The same is true for $P_{\mu\tau}$ in Figure~\ref{fig:Pmutau-corr}, although the difference between the corrections up to the second order and up to the third order is overshadowed by the correction up to the zeroth order, which is notably different than the numerical solution. We note that including higher-order corrections into the series expansions yield more accurate results, as more oscillatory frequencies are taken into consideration.

% CORRECTIONS FROM DIFFERENT ORDERS
\begin{figure}[!t]
    \centering
    \begin{subfigure}{.48\linewidth}
    \centering
    \includegraphics[width=1.\linewidth]{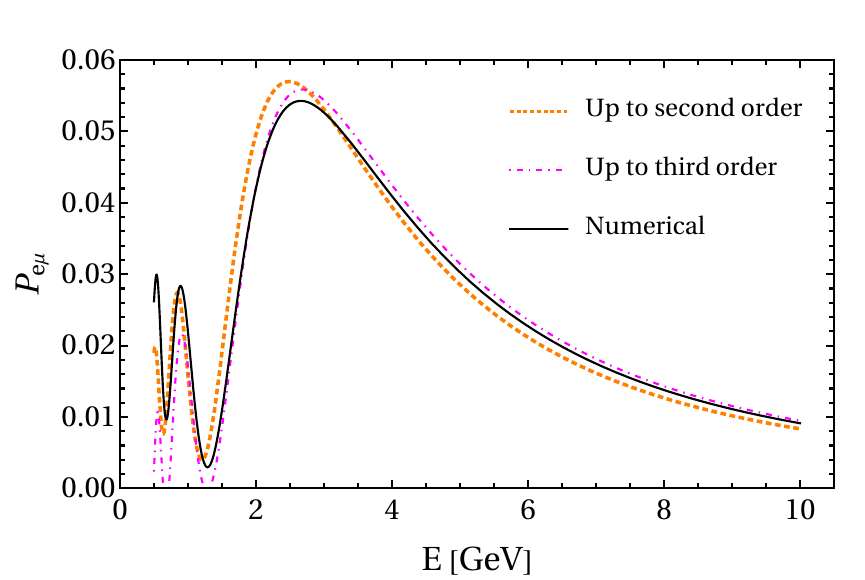}
    \caption{\label{fig:Pemu-corr}Contributions for $P_{e\mu}$.}
    \end{subfigure}
    \begin{subfigure}{.48\linewidth}
    \centering
    \includegraphics[width=1.\linewidth]{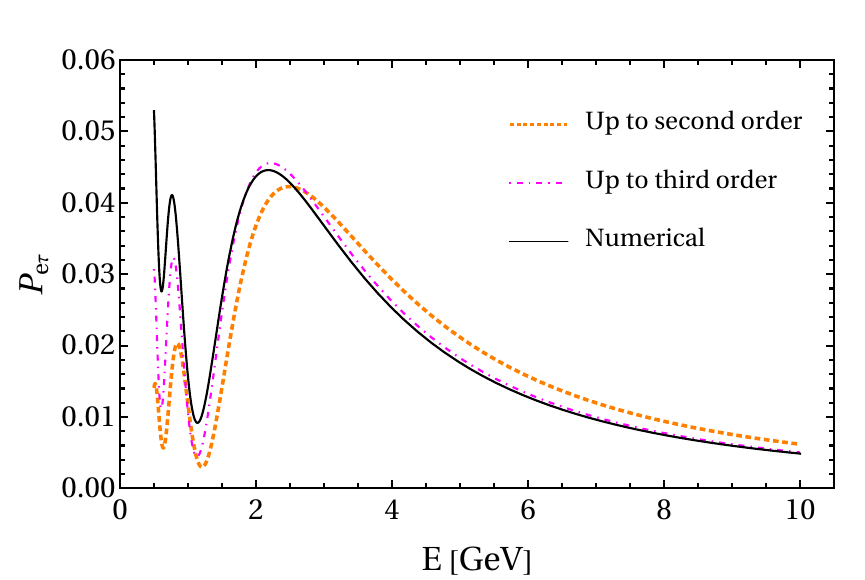}
    \caption{\label{fig:Petau-corr}Contributions for $P_{e\tau}$.}
    \end{subfigure}\\
    \begin{subfigure}{.48\linewidth}
    \centering
    \includegraphics[width=1.\linewidth]{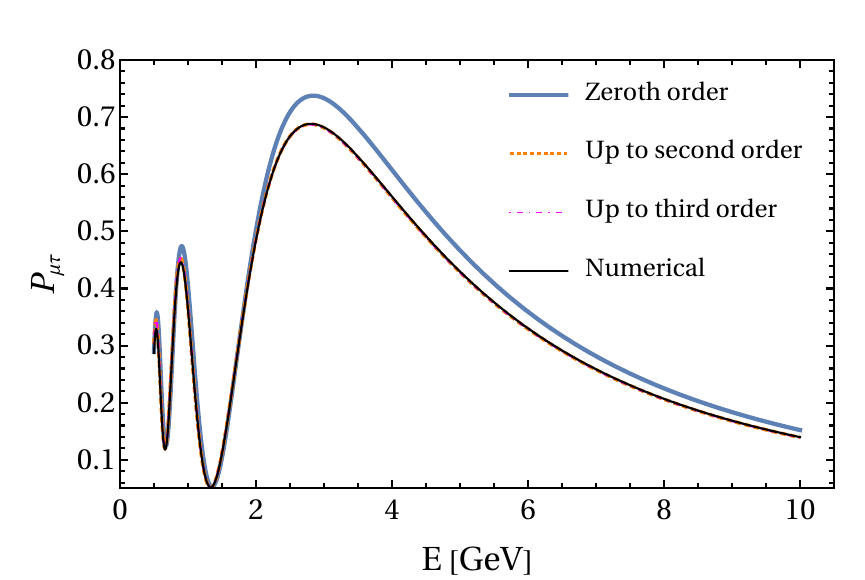}
    \caption{\label{fig:Pmutau-corr}Contributions for $P_{\mu\tau}$.}
    \end{subfigure}\\    
    \caption{The contributions to the neutrino oscillation probabilities from the various orders of $\lambda$ in the series expansion approach, as well as the numerically calculated probabilities. The neutrino oscillation probabilities are plotted up to the zeroth, second and third orders in $\lambda$ for (a) $P_{e\mu}$, (b) $P_{e\tau}$ and (c) $P_{\mu\tau}$, respectively, assuming the baseline length $L = 1300$~km.}
    \label{fig:Corrections}
\end{figure}

% ERROR PLOTS
%\textbf{Errors}
\begin{figure}[!t]
  \centering
  \begin{subfigure}{.43\linewidth}
    \centering
    \includegraphics[width = 1.0\linewidth]{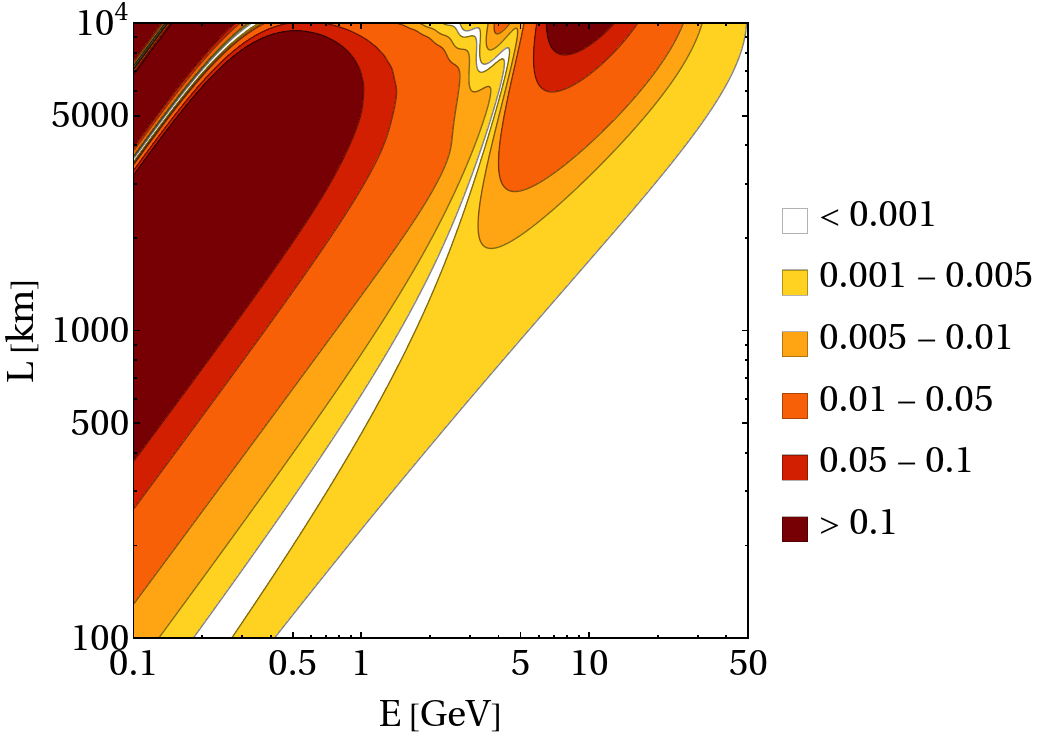}
    \caption{Absolute error $|\Delta P_{ee}|$.}
  \end{subfigure}%
  \hspace{2em}% Space between image A and B
  \begin{subfigure}{.43\linewidth}
    \centering
    \includegraphics[width = 1.0\linewidth]{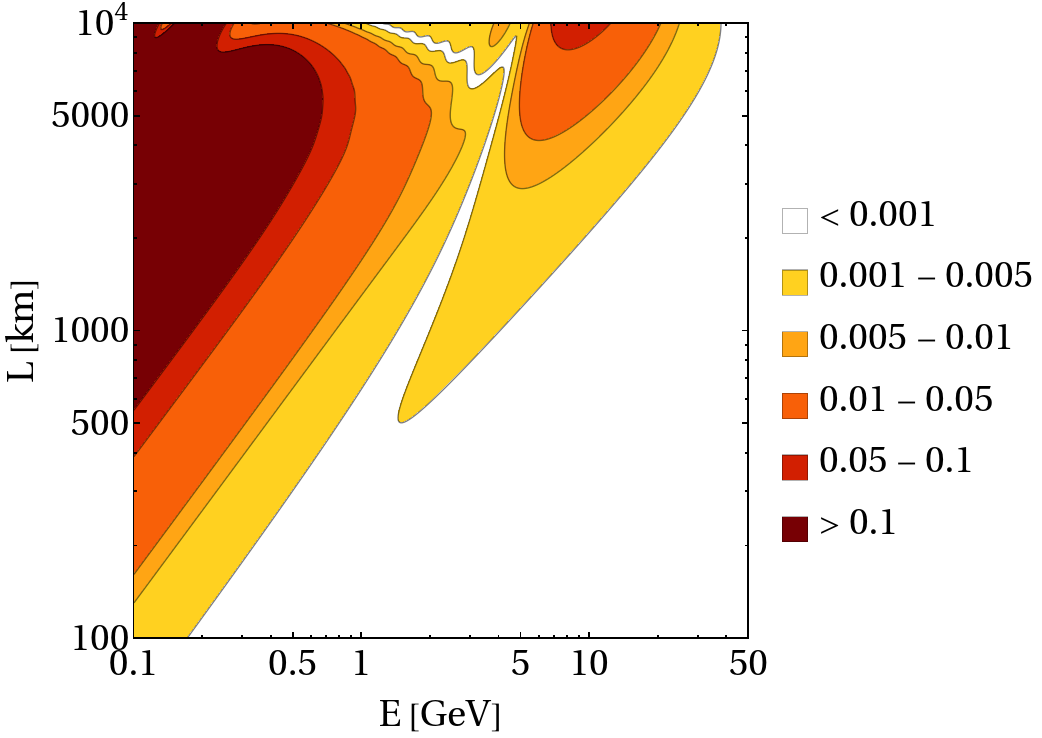}
    \caption{Absolute error $|\Delta P_{e\mu}|$.}
  \end{subfigure}\\[3ex]
    \begin{subfigure}{.43\linewidth}
    \centering
    \includegraphics[width = 1.0\linewidth]{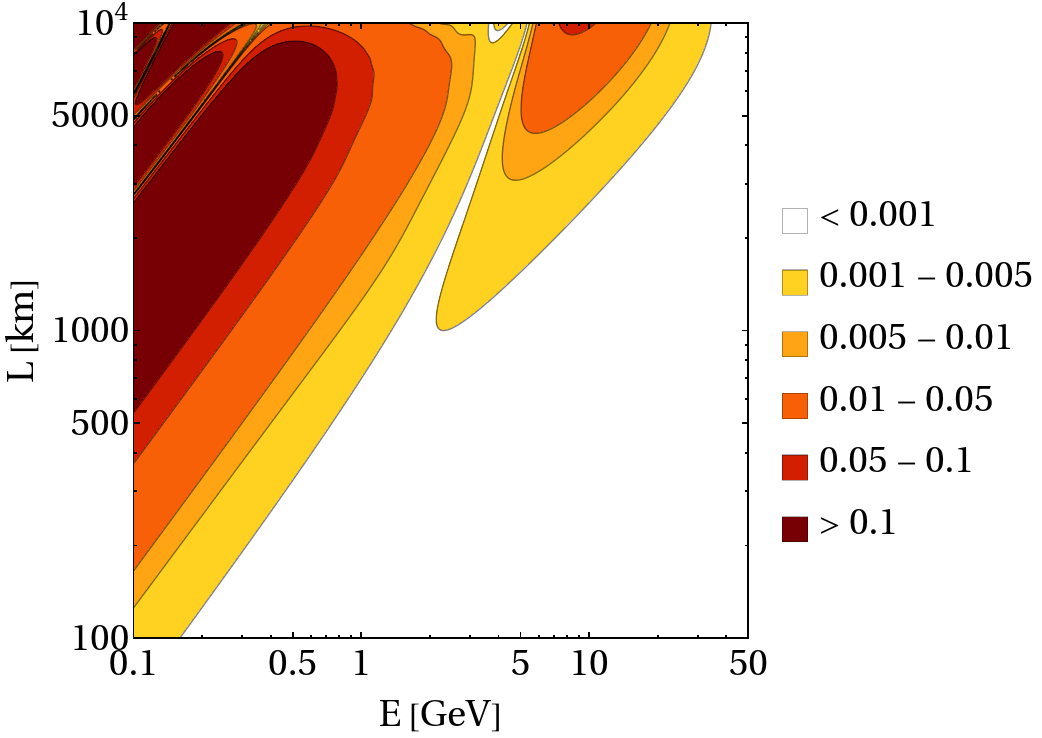}
    \caption{Absolute error $|\Delta P_{e\tau}|$.}
  \end{subfigure}
    \hspace{2em}% Space between image A and B
    \begin{subfigure}{.43\linewidth}
    \centering
    \includegraphics[width = 1.0\linewidth]{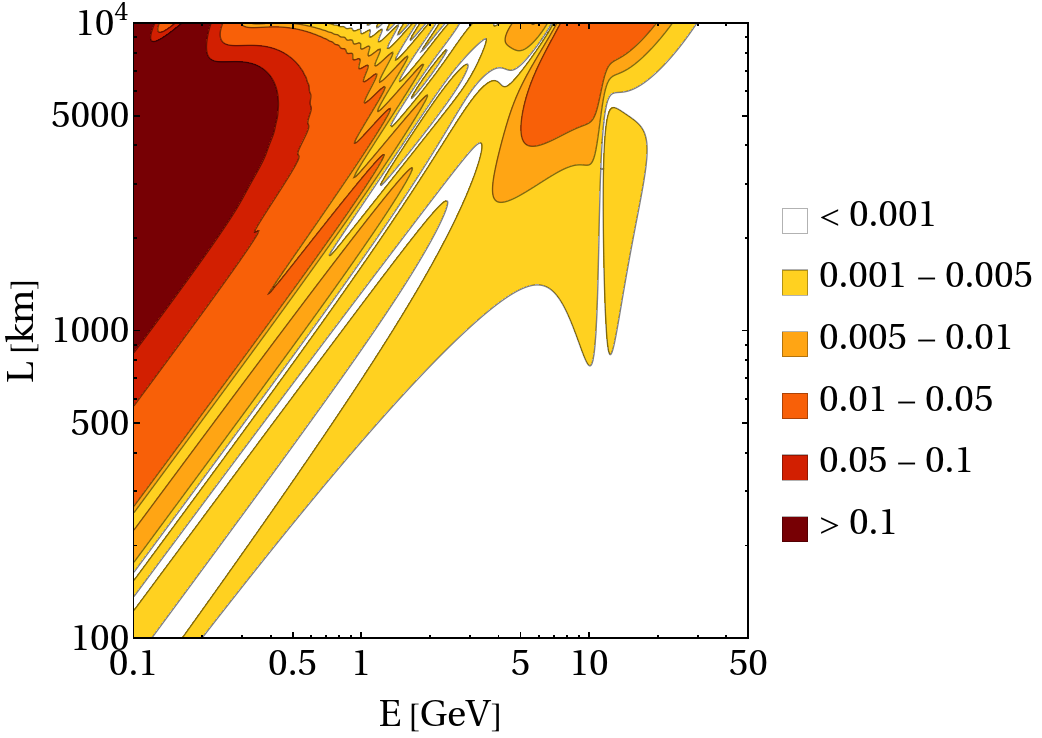}
    \caption{Absolute error $|\Delta P_{\mu\mu}|$.}
  \end{subfigure}\\[3ex]
      \begin{subfigure}{.43\linewidth}
    \centering
    \includegraphics[width = 1.0\linewidth]{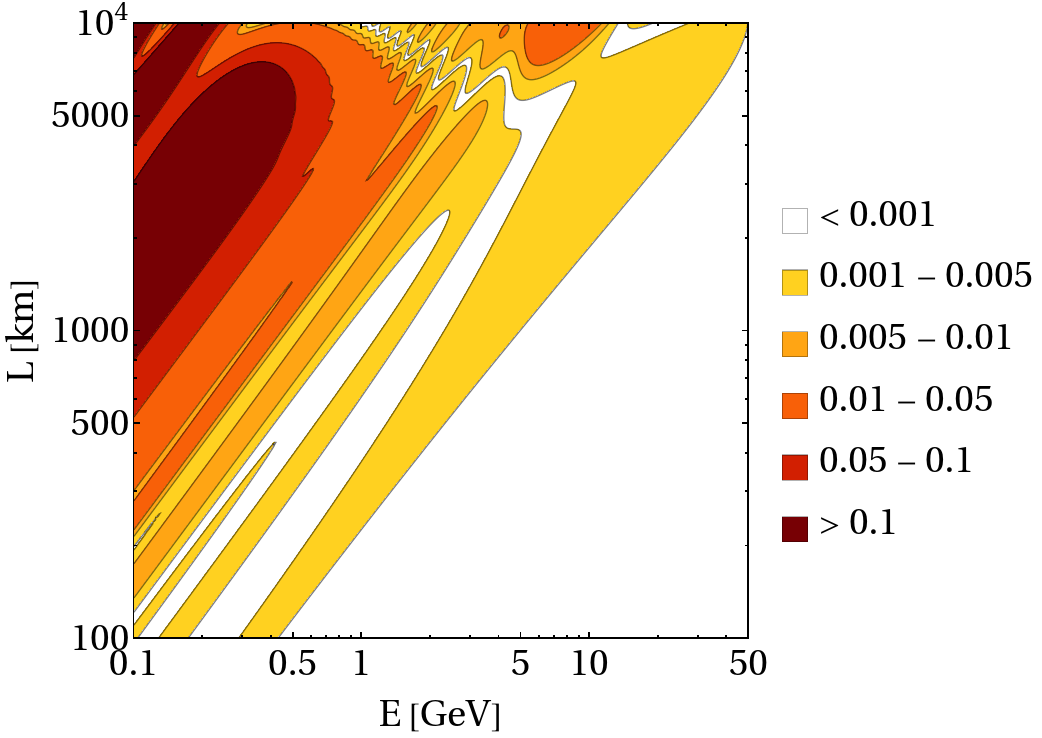}
    \caption{Absolute error $|\Delta P_{\mu\tau}|$.}
  \end{subfigure}
    \hspace{2em}% Space between image A and B
    \begin{subfigure}{.43\linewidth}
    \centering
    \includegraphics[width = 1.0\linewidth]{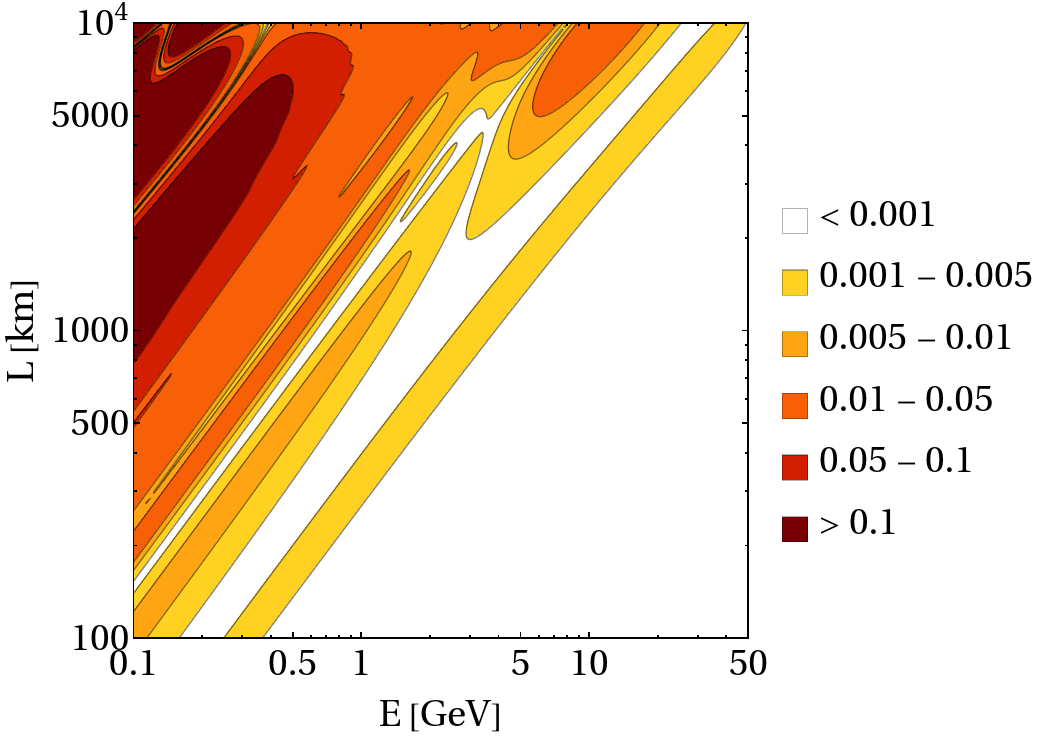}
    \caption{Absolute error $|\Delta P_{\tau\tau}|$.}
  \end{subfigure}\\[2ex]
  \caption{
  Absolute differences for numerically computed probabilities and probabilities obtained from the series expansions. The differences are shown for probabilities (a) $P_{ee}$, (b) $P_{e\mu}$, (c) $P_{e\tau}$, (d) $P_{\mu\mu}$, (e) $P_{\mu\tau}$ and (f) $P_{\tau\tau}$ assuming neutrino energies $E = 0.1$~GeV--$50$~GeV and baseline lengths $L = 100$~km--$10~000$~km.}
  \label{fig:Errors}
\end{figure}

The absolute errors $\left|\Delta P_{\alpha\beta}\right|\equiv\big|P_{\alpha\beta}^\text{expansion}-P_{\alpha\beta}^\text{numerical}\big|$ for $\alpha, \beta = e, \mu, \tau$ are further illustrated with contour plots in Figure~\ref{fig:Errors}. The figure shows the differences between the approximate and numerically calculated probabilities in logarithmic scale for neutrino energies $E = 0.1$~GeV--$50$~GeV and baseline lengths $L = 100$~km--$10~000$~km. As is shown in the legend bar, regions with darker shades correspond to smaller absolute errors and those with lighter shades to larger errors, respectively. We observe that the absolute errors are the smallest in the lower-right triangular half-plane, where the errors are of order $10^{-3}$. The series expansion method appears to be the most accurate in regions where neutrino energies are high and baseline lengths are short. The absolute errors then grow as neutrino energies become lower and baseline lengths get longer. The absolute errors also display oscillatory features, as there are regions where the approximate probabilities are relatively close to the numerical ones despite the accuracy condition~(\ref{Accuracy condition}) no longer being valid. For those regions, the neutrino oscillations driven by $\Delta m_{21}^2 L / (4 E)$, for example, can be expected to have non-negligible effects on the neutrino oscillation probabilities. In that particular case, the probabilities are altered by $\Delta P_{\alpha \beta} \simeq (1/2)\sin^2 2\theta_{12} \sin^2 (\alpha \Delta)$, which yields $\Delta P_{\alpha \beta} \approx 0.007 \sim \lambda^3$ for $E = 1$~GeV and $L = 1300$~km. This contribution could also explain the gap between the exact result and the series expansion prediction in Figure~\ref{fig:2b}. Having analytical expressions for the neutrino oscillation probabilities is useful even at an approximate level, as they demonstrate the functional dependence on the neutrino oscillation parameters. It is also important to know the limitations of the approximate probabilities for different baseline lengths and neutrino energies. In the rest of this section, we examine the relevance of the series expansions in greater detail. 

In Figure~\ref{fig:Pemu}, we show the transition probability $P_{e\mu}$ and the absolute error $\Delta P_{e\mu}$ for neutrino energies $E = 0.15$~GeV--$3$~GeV and the baseline length $L = 295$~km. This configuration corresponds to the experiment setup of T2K~\cite{T2K:2013ppw}, which is a long-baseline neutrino oscillation experiment currently taking data in Japan.  The series expansion for the conversion probability $P_{\mu e}$ can be obtained from equation~(\ref{Pemu}) through the transformation $\delta \rightarrow -\delta$. As before, we take the decay parameter to be $\gamma = 0.1$ and the average matter density $\rho = 3$~g/cm$^3$. The absolute error is computed as the difference of the numerically computed probability $P_{e\mu}$ and the matching series expansion formula. We observe in Figure~\ref{fig:Pemu} that the analytical and numerically calculated probabilities are in good agreement with neutrino energies above $0.4$~GeV. Figure~\ref{fig:Pemu:a} shows that the locations of maxima and minima are predicted correctly by the series expansion formulas, with the first oscillation maximum taking place at about $0.65$~GeV and the first minimum above $3$~GeV. The absolute errors in Figure~\ref{fig:Pemu:b} indicate that the absolute error is the most significant at $0.15$~GeV, where the error stays below $0.02$. We remark that the absolute error has its its minimum at $0.6$~GeV, where the error is of order $10^{-5}$. At higher energies, the absolute error grows to about $10^{-3}$ and starts falling again. The approximate probabilities are therefore most suitable to study probabilities at $0.6$~GeV, which also coincides with the peak energy of the T2K setup.

\begin{figure}[!t]
  \centering
  \begin{subfigure}{.485\linewidth}
    \centering
    \includegraphics[width = 1.\linewidth]{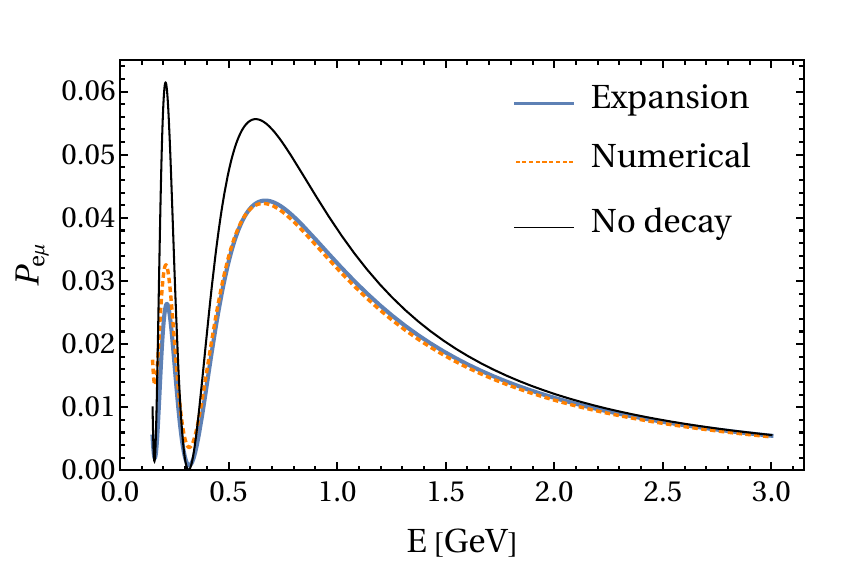}
    \caption{\label{fig:Pemu:a}Transition probability $P_{e\mu}$.}
  \end{subfigure}%
  \hspace{0.8em}% Space between image A and B
  \begin{subfigure}{.485\linewidth}
    \centering
    \includegraphics[width = 1.\linewidth]{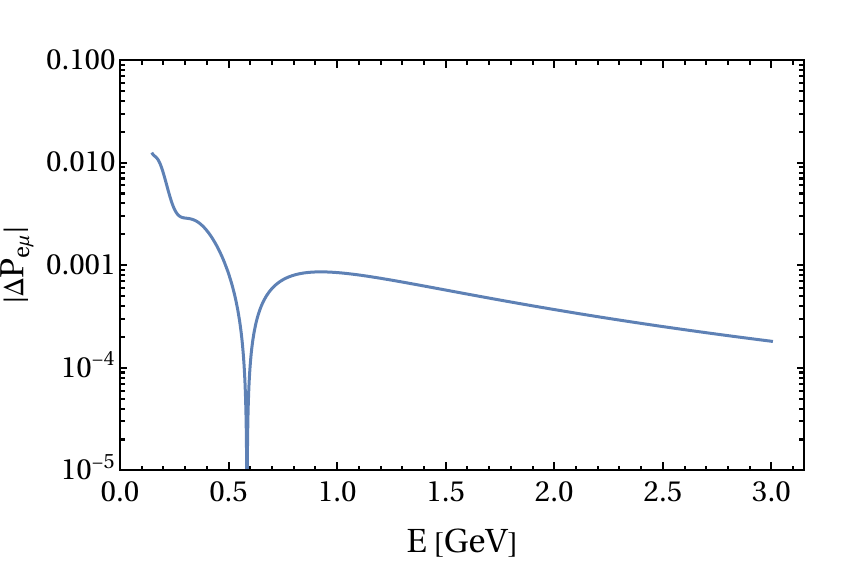}
    \caption{\label{fig:Pemu:b}Approximation error $|\Delta P_{e\mu}|$.}
  \end{subfigure}
  \caption{The transition probability $P_{e\mu}$ and the absolute error $|\Delta P_{e\mu}|$, plotted in the energy range $E = 0.15$~GeV--$3$~GeV with baseline length $L=295$~km, corresponding to the conditions at the T2K setup.}
  \label{fig:Pemu}
\end{figure}

We next investigate the accuracy of the series expansion that has been derived for the muon neutrino survival probability $P_{\mu\mu}$. The performance of the series expansion is shown in Figure~\ref{fig:Pmumu}, where the probability is plotted with both the approximate method and numerical calculation. The approximate and numerically acquired probabilities are shown with the blue solid and yellow dashed lines, respectively. The approximate probabilities are computed using the series expansion presented in equations~(\ref{Pmumu0})--(\ref{Pmumu3}). The probabilities are plotted for neutrino energies $E = 0.5$~GeV--$10$~GeV and the baseline length $L = 1300$~km. The setup therefore matches with the DUNE setup~\cite{DUNE:2015lol}, which is a neutrino oscillation experiment currently under construction in USA. Figure~\ref{fig:Pmumu:a} shows that the series expansion approximates the numerically computed probabilities well, while Figure~\ref{fig:Pmumu:b} confirms that the absolute error $|\Delta P_{\mu\mu}|$ stays below $10^{-2}$ for all considered neutrino energies and below $3\times10^{-3}$ for neutrino energies higher than $0.75$~GeV. We note that in the absence of neutrino decay, the probability corresponding to $\nu_\mu \rightarrow \nu_\mu$ channel varies between zero and unity in the energies considered in Figure~\ref{fig:Pmumu}. The effect of neutrino decay is therefore significant for $P_{\mu\mu}$ for the chosen value $\gamma = 0.1$. In comparison, the effect appears to be less noticeable in the conversion probability $P_{e\mu}$. Many high-energy accelerator neutrino experiments like DUNE use muon neutrino beams and are therefore suitable for the search of invisible neutrino decay.

The accuracy of the series expansions of $P_{\mu e}$ and $P_{\mu \mu}$ have previously been studied in Ref.~\cite{Chattopadhyay:2022ftv}, where the probabilities have been computed for the baseline length of the DUNE setup, $L = 1300$~km, and the magic baseline, $L = 7000$~km. We note that our results in Figure~\ref{fig:Pemu} are mostly in agreement with the probabilities that were provided in Ref.~\cite{Chattopadhyay:2022ftv}. The main difference comes from the absolute error, which is found to be an order of magnitude smaller for the analytical formulas that have been obtained in the present work. In the energy range $0.75$~GeV--$3$~GeV, in particular, the magnitude of the absolute error in Figure~\ref{fig:Pmumu:b} is of order $10^{-3}$, whereas the magnitude of the corresponding error in Ref.~\cite{Chattopadhyay:2022ftv} is of order $10^{-2}$. This difference in the accuracy arises from the order of the series expansion used in both works, which is $\mathcal{O}(\lambda^3)$ in our work and $\mathcal{O}(\lambda^2)$ in Ref.~\cite{Chattopadhyay:2022ftv}.

We finally consider the tau neutrino appearance probability $P_{\mu\tau}$. Figure~\ref{fig:Pmutau-OPERA} shows the probability and its absolute error in the conditions of the OPERA setup~\cite{Acquafredda:2009zz}, which entails neutrino energies $E = 0.3$~GeV--$8$~GeV and the baseline length $L = 730$~km. This energy range includes the first three oscillation maxima and two minima of $P_{\mu\tau}$. As before, we have plotted the probability assuming the decay parameter $\gamma = 0.1$ and matter density $\rho = 3$~g/cm$^3$. The probabilities and their errors look qualitatively similar to those plotted for $P_{\mu\mu}$ in Figure~\ref{fig:Pmumu}. The maximum value of the absolute error $\Delta P_{\mu\tau} = 0.02$ is found at the lowest considered energy $E = 0.3$~GeV, while neutrino energies higher than $0.5$~GeV yield errors that are of order $10^{-3}$. There are three local minima in the absolute error curve, with the lowest values being of order 10$^{-5}$.

\begin{figure}[t]
  \centering
  \begin{subfigure}{.485\linewidth}
    \centering
    \includegraphics[width = 1.\linewidth]{Figures/Pmumu-scaled2.pdf}
    \caption{\label{fig:Pmumu:a}Transition probability $P_{\mu\mu}$.}
  \end{subfigure}%
  \hspace{0.8em}% Space between image A and B
  \begin{subfigure}{.485\linewidth}
    \centering
    \includegraphics[width = 1.\linewidth]{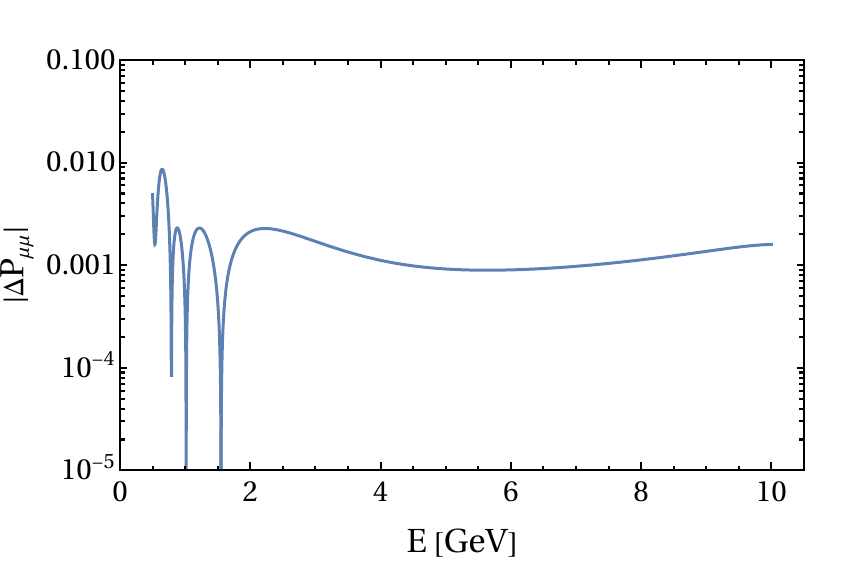}
    \caption{\label{fig:Pmumu:b}Approximation error $|\Delta P_{\mu\mu}|$.}
  \end{subfigure}
  \caption{The survival probability $P_{\mu\mu}$ and the absolute error $|\Delta P_{\mu\mu}|$, plotted in the energy range $E = 0.5$~GeV--$10$~GeV with baseline length $L=1300$ km, corresponding to the conditions of the DUNE setup.}
  \label{fig:Pmumu}
\end{figure}

\begin{figure}[!t]
  \centering
  \begin{subfigure}{.485\linewidth}
    \centering
    \includegraphics[width = 1.\linewidth]{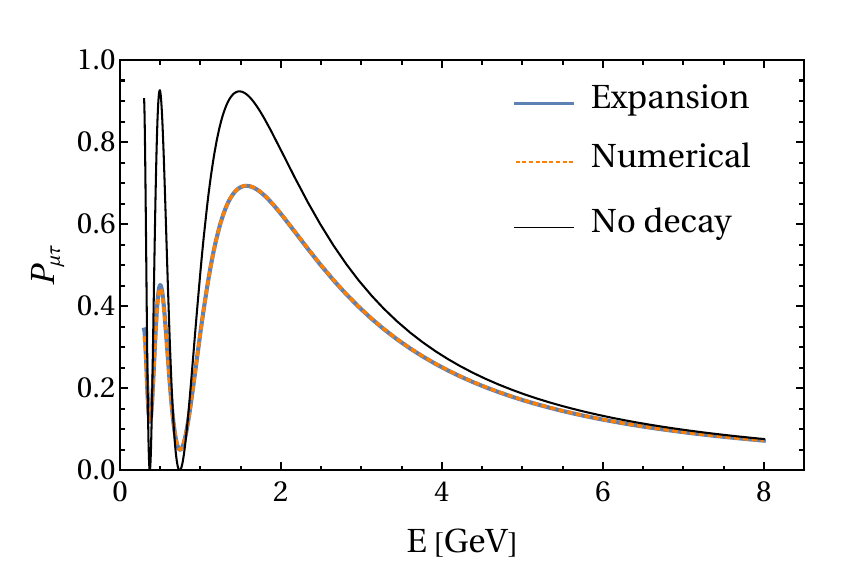}
    \caption{Transition probability $P_{\mu\tau}$.}
  \end{subfigure}%
  \hspace{0.8em}% Space between image A and B
  \begin{subfigure}{.485\linewidth}
    \centering
    \includegraphics[width = 1.\linewidth]{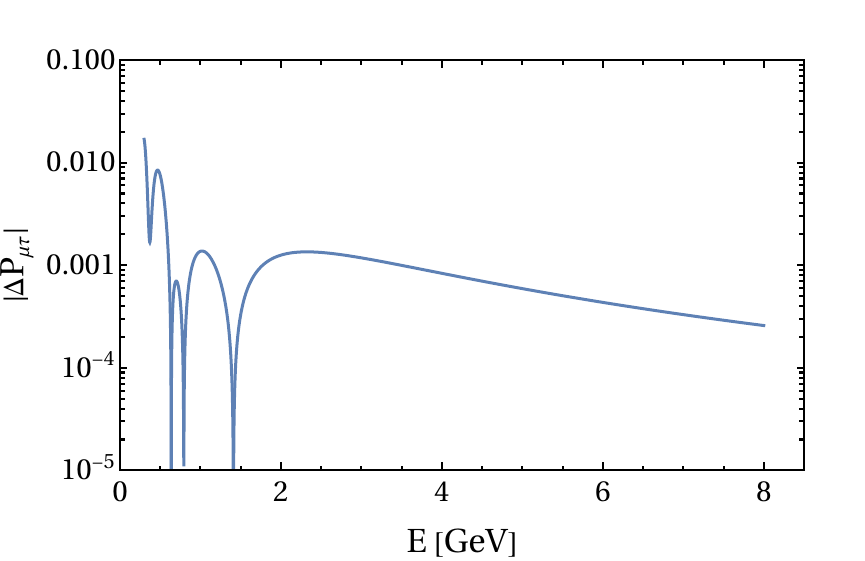}
    \caption{Approximation error $|\Delta P_{\mu\tau}|$.}
  \end{subfigure}
  \caption{The transition probability $P_{\mu\tau}$ and the absolute error $|\Delta P_{\mu\tau}|$, plotted in the energy range $E = 0.3$~GeV--$8$~GeV with baseline length $730$~km, corresponding to the conditions of the OPERA setup.}
  \label{fig:Pmutau-OPERA}
\end{figure}

In addition to the scenarios that are studied in this section, the probability formulas that have been derived in this work are applicable to a wide range of neutrino experiments. Long-baseline experiments including DUNE and T2HK~\cite{Abe:2015zbg} benefit not only from having compact analytical probability formulas for the muon neutrino survival and conversion probabilities $P_{\mu\mu}$ and $P_{\mu e}$, but understanding their data also requires knowledge on the electron neutrino probability $P_{e e}$. Experiments with high-energy neutrino beams like DUNE are furthermore expected to have limited sensitivity to tau neutrino appearance~\cite{DeGouvea:2019kea} and therefore benefit from the series expansions of $P_{\mu \tau}$ and $P_{e \tau}$. The tau neutrino survival probability $P_{\tau \tau}$ is not directly testable in neutrino experiments, but it can be indirectly measured through unitarity relations~\cite{Martinez-Soler:2021sir}, which have to be modified to account for neutrino decay. We note that our results are not applicable to conditions where the accuracy relation~(\ref{Accuracy condition}) is not valid or the matter density profile cannot be approximated with a constant density. The probability formulas presented in this work are therefore normally not suitable to describe neutrino oscillations in solar neutrino experiments like SNO~\cite{SNO:2001kpb} and BOREXINO~\cite{BOREXINO:2023asq}, reactor neutrino experiments like Daya Bay~\cite{DayaBay:2016ggj} and JUNO~\cite{An:2015jdp}, see also Ref.~\cite{Porto-Silva:2020gma}, or atmospheric neutrino experiments like IceCube~\cite{IceCube:2019dqi}.

\section{\label{sec:summary}Summary and conclusions}
In this paper, we have presented complete set of approximate formulas for neutrino oscillation probabilities in presence of invisible neutrino decay. We have derived compact analytical formulas for the oscillation channels $\nu_e \rightarrow \nu_e$, $\nu_e \rightarrow \nu_\mu$, $\nu_e \rightarrow \nu_\tau$, $\nu_\mu \rightarrow \nu_\mu$ and $\nu_\mu \rightarrow \nu_\tau$ by taking series expansions in small parameters $s_{13}$ and $\alpha$. Our results include matter effects with constant density. The probability formulas have been obtained using the Cayley--Hamilton formalism and perturbative diagonalization. Expressing results in terms of $\lambda =$ 0.2, the analytical probabilities have been derived up to $\mathcal{O}(\lambda^2)$ for the probability $P_{ee}$ and $\mathcal{O}(\lambda^3)$ for the probabilities $P_{e\mu}$, $P_{e\tau}$, $P_{\mu\mu}$ and $P_{\mu\tau}$, respectively. We have additionally studied the accuracy of the analytical probability formulas for a wide range of neutrino energies $E$ and baseline lengths $L$ and examined the impact on unitarity relations. Below we summarize the main findings of this work:

\begin{itemize}
    \item The probability formulas derived are valid for the accuracy condition $\alpha \Delta = \Delta m_{21}^2 L / (4E) \ll 1$, which holds up to $L/E \ll 10^4$~km/GeV. The absolute errors in this region are generally of order $10^{-3}$, indicating a good agreement between numerically computed probabilities and the analytical probability formulas.
    \item The neutrino oscillation probabilities that are obtained with the series expansions for $\gamma = 0.1$ are more consistent with the probabilities that are calculated numerically with neutrino decay than the ones that are obtained without decay. This is shown in Figures~\ref{fig:Probabilities}, \ref{fig:Pemu}, \ref{fig:Pmumu} and \ref{fig:Pmutau-OPERA}.
    \item The accuracy of the derived probability formulas is particularly good for the conditions of T2K and T2HK, where the absolute errors for $P_{e\mu}$ are of order $10^{-5}$ for neutrino energies $E \simeq 0.6$~GeV. We also found a good agreement for the DUNE setup, where the absolute error pertaining to $P_{\mu\mu}$ is of order $10^{-2}$ at the lowest neutrino energies and stays under $3\times10^{-3}$ for energies higher than $0.75$~GeV.
    \item Neutrino oscillation probabilities with neutrino decay have previously been studied with the series expansion method in Ref.~\cite{Chattopadhyay:2022ftv}. We find that our results for the probabilities $P_{e\mu}$ and $P_{\mu\mu}$ are in agreement with the equivalent relations in Ref.~\cite{Chattopadhyay:2022ftv}. Our work reproduces the formula for $P_{e\mu}$ completely and extends the formula for $P_{\mu\mu}$ to include terms up to $\mathcal{O}(\lambda^3)$, respectively.
    \item Probabilities without neutrino decay have been studied using the series expansion method in Ref.~\cite{Akhmedov:2004ny}. Comparing the probability formulas derived at the no-decay limit $\gamma \rightarrow 0$ in this work and the corresponding formulas in Ref.~\cite{Akhmedov:2004ny}, we find our formulas for $P_{ee}$, $P_{e\mu}$ and $P_{e\tau}$ to be consistent with those in Ref.~\cite{Akhmedov:2004ny}. We also find a third-order term in the formulas for $P_{\mu\mu}$ and $P_{\mu\tau}$ that is different between Ref.~\cite{Akhmedov:2004ny} and this work. The term is also present in $P_{\tau\tau}$, which can be obtained from $P_{\mu\mu}$~\cite{Akhmedov:2004ny}. We estimate the difference to be of order $10^{-3}$.
    \item Probability formulas for calculating neutrino oscillation probabilities in vacuum have also been studied in the OMSD limit~\cite{Choubey:2020dhw}. We find the probability formulas derived in this work to be in agreement with those in Ref.~\cite{Choubey:2020dhw} after applying the vacuum limit $A\rightarrow 0$.
    \item Unitarity is generally violated in presence of neutrino decay. In the muon and tau neutrino probabilities, we find the violation to be proportional to $1-\text{e}^{-4 \gamma \Delta}$ at the leading order, whereas in the electron neutrino probabilities the violation is of the order $\mathcal{O}(\lambda^2)$.
\end{itemize}

In conclusion, we have presented series expansions to approximate the probabilities for neutrino oscillations in constant matter density and one decaying neutrino mass eigenstate. The formulas derived in this work are in agreement with those provided in Ref.~\cite{Chattopadhyay:2022ftv} for $P_{e\mu}$ and $P_{\mu\mu}$, whereas the rest of our results are novel. We have provided a detailed study on the numerical accuracy of the formulas and discussed their relevance for various neutrino experiments.

\appendix

\section{\label{sec:appA}Matrix exponential in the Cayley--Hamilton formalism}

In this section, we briefly describe the Cayley--Hamilton formalism. In order to obtain the exponent form of an $N \times N$ matrix $\mathcal{M}$, one can write the series
\begin{equation}
\text{e}^\mathcal{M} = \sum_{n=0}^{\infty} \frac{1}{n!} \mathcal{M}^n.
\label{eq:appA1}
\end{equation}
Following from linear algebra, the characteristic equation of $\mathcal{M}$ is
\begin{equation}
\chi_\mathcal{M}(\lambda) \equiv \det(\mathcal{M}-\lambda \boldsymbol{I}) = 0,
\label{eq:appA2}
\end{equation}
where $\chi_\mathcal{M}$ is the characteristic polynomial, $\boldsymbol{I}$ is the $N \times N$ identity matrix and $\lambda$ is an eigenvalue of $\mathcal{M}$. The characteristic polynomial of $\mathcal{M}$ can be written in terms of its eigenvalues as
\begin{equation}
\chi_\mathcal{M}(\lambda) = \lambda^N + a_{N-1} \lambda^{N-1} + \ldots + a_1 \lambda + a_0 = 0,
\label{eq:appA3}
\end{equation}
where $a_0, a_1, \ldots, a_{N-1}$ are some coefficients.

The Cayley--Hamilton theorem states that every matrix $\mathcal{M}$ satisfies the characteristic equation $\chi_\mathcal{M} = 0$, that is,
\begin{equation}
\mathcal{M}^N = -a_{N-1} \mathcal{M}^{N-1} - \ldots - a_1 \mathcal{M} - a_0 \boldsymbol{I},
\label{eq:appA4}
\end{equation}
where the highest power of $\mathcal{M}$ is written in terms of its lower powers. Therefore, for any $p \geq N$, there exist coefficients $b_0, b_1, \ldots, b_{N-1}$ that satisfy the equation
\begin{equation}
M^p = b_{N-1} \mathcal{M}^{N-1} + \ldots + b_1 \mathcal{M} + b_0 \boldsymbol{I}.
\label{eq:appA5}
\end{equation}
The Cayley--Hamilton formalism makes it possible to reduce the infinite series of exponential matrix $\text{e}^\mathcal{M}$ to a finite one, using a set of coefficients $c_0, c_1, \ldots, c_{N-1}$ as
\begin{equation}
\text{e}^\mathcal{M} = c_0 \boldsymbol{I} + c_1 \mathcal{M} + \ldots + c_{N-1} \mathcal{M}^{N-1} = \sum_{n=0}^{N-1} c_n \mathcal{M}^n,
\label{eq:appA6}
\end{equation}
where the coefficients $c_i$ are functions of the coefficients $a_i$. In neutrino oscillations, the formalism is useful in expressing the time-evolution operator $S = \text{e}^{-i\mathcal{H}L}$ with a reduced set of operators. For three active flavor states, the reduction can be performed in three terms.

\section{\label{sec:appB}Perturbative diagonalization}
In this work, we adopt the approach in Ref.~\cite{Akhmedov:2004ny} for the perturbative diagonalization of the neutrino mixing matrix. Let us consider a $3\times3$ matrix $M$ that can be diagonalized by another matrix $W$:
\begin{equation}
M = W \hat{M} W^\dagger,
\label{eq:appB1}
\end{equation}
where $\hat{M} = {\rm diag}(\lambda_1, \lambda_2, \lambda_3)$ is the diagonalized matrix and $\lambda_1$, $\lambda_2$ and $\lambda_3$ are eigenvalues of $M$. Correspondingly, one may obtain the eigenvectors $v_1$, $v_2$ and $v_3$ from the columns of the diagonalizing matrix $W$. In the perturbative approach, the quantities can be decomposed into contributions consisting of terms in various orders of perturbation parameters, such as $M = M^{(0)} + M^{(1)} + M^{(2)} + \ldots$ . This method can be applied to both eigenvalues and eigenvectors. In three-flavor oscillations, a convenient choice for $M$ is
\begin{equation}
M \equiv U_{13} U_{12} {\rm diag}(0, \alpha, 1 - i \gamma) U_{12}^\dagger U_{13}^\dagger + {\rm diag}(A, 0, 0),
\label{eq:appB2}
\end{equation}
where $U_{ij}$ are elements of the Pontecorvo--Maki--Nakagawa--Sakata matrix and $A$ represents the matter potential. The Hamiltonian then takes the well-known form
\begin{equation}
\mathcal{H} = \frac{\Delta m_{31}^2}{2 E} U_{23} U_\delta M U_\delta^\dagger U_{23}^\dagger.
\label{eq:appB3}
\end{equation}
This turns out to be a convenient choice, as the matter potential matrix is unaffected by the action of $U_{23}$ and $U_\delta$. The matrix product $U_{23} U_\delta W \equiv \tilde{U}$ therefore represents the leptonic mixing matrix in matter. Writing equation~(\ref{eq:appB2}) in full form and and grouping its terms in orders of the expansion parameters $s_{13}$ and $\alpha$, we obtain
\begin{align}
M^{(0)} &= \begin{pmatrix}A & 0 & 0\\0 & 0 & 0\\0 & 0 & 1 -i\gamma\end{pmatrix},\\
M^{(1)} &= \begin{pmatrix}0 & 0 & c_{13} s_{13} (1-i\gamma)\\0 & 0 & 0\\c_{13} s_{13} (1-i\gamma) & 0 & 0\end{pmatrix},\\
M^{(2)} &= \begin{pmatrix}\alpha s_{12}^2 + s_{13}^2 (1-i\gamma) & \alpha c_{13} c_{12} s_{12} & 0\\\alpha c_{13} c_{12} s_{12} & \alpha c_{12}^2 & 0\\0 & 0 & -s_{13}^2 (1-i\gamma)\end{pmatrix}.
\label{eq:appB4}
\end{align}
Since $M^{(0)}$ is diagonal, the zeroth-order contributions to the eigenvalues are diagonal elements, whereas the corresponding eigenvectors are the standard Euclidean basis vectors $e_1$, $e_2$ and $e_3$. The first and second order corrections to the eigenvalues on the other hand are given as
\begin{align}
\lambda_i^{(1)} &= M_{ii}^{(1)},\\
\lambda_i^{(2)} &= M_{ii}^{(2)} + \sum_{j \neq i} \frac{\left( M_{ij}^{(1)} \right)^2}{\lambda_i^{(0)} - \lambda_j^{(0)}},
\label{eq:appB5}
\end{align}
and to the eigenvectors as
\begin{align}
v_i^{(1)} &= \sum_{j \neq i} \frac{M_{ij}^{(1)}}{\lambda_i^{(0)} - \lambda_j^{(0)}} e_j,\\
v_i^{(2)} &= \sum_{j \neq i} \frac{1}{\lambda_i^{(0)} - \lambda_j^{(0)}} \left[ M_{ij}^{(2)} + \left( M^{(1)} v_i^{(1)} \right)_j - \lambda_i \left( v_i^{(1)} \right)_j \right] e_j,
\label{eq:appB6}
\end{align}
respectively. The energy eigenvalues are therefore written for the three-flavor case as
\begin{equation}
E_i = \frac{\Delta m_{31}^2}{2E} \lambda_i,
\label{eq:appB7}
\end{equation}
where $i =$ 1, 2 and 3.

\bibliography{references}
\end{document}